%% file: PaperDraft.tex
\documentclass[journal,twocolumn]{IEEEtran}

\usepackage{cite}
\usepackage[colorlinks=true,linkcolor=blue,allcolors=blue]{hyperref}%

\input{preambleT.tex}
\usepackage{lipsum}
\usepackage{mathtools}
\usepackage{cuted}
\usepackage{physics}

\begin{document}
%
\title{Flood Risk Mitigation and Valve Control in Stormwater Systems: State-Space Modeling, Control Algorithms, and Case Studies}
%
%
%

\author{{Marcus N. Gomes J\'{u}nior\textsuperscript{$\dag,\ast$}, Marcio H. Giacomoni\textsuperscript{$\ddag$}, Ahmad F. Taha\textsuperscript{$\dagger\dagger$}, and Eduardo M. Mendiondo\textsuperscript{\S}}
\thanks{\dag Department of Civil and Environmental Engineering, The University of Texas at San Antonio, One UTSA Circle, BSE 1.310, TX 78249 (marcusnobrega.engcivil@gmail.com). 

$\ddag$Department of Civil and Environmental Engineering, The University of Texas at San Antonio, One UTSA Circle, BSE 1.346, TX 78249 (marcio.giacomoni@utsa.edu).

$\dagger\dagger$Department of Civil and Environmental Engineering, Vanderbilt University Jacobs Hall, Office $\#$ 293, 24\textsuperscript{th} Avenue South, Nashville, TN 37235 (ahmad.taha@vanderbilt.edu).

\textsuperscript{\S}Department of Hydraulic Engineering and Sanitation, University of Sao Paulo, Sao Carlos School of Engineering, 13566-590 (emm@sc.usp.br).

*Corresponding author.

This work was financially supported by the City of San Antonio and by the San Antonio River Authority. This work is also partially supported by National Science Foundation (NSF) under Grant 2015671.}}

	 \markboth{Journal of Water Resources Planning and Management, In Press, May 2022}{Gomes J\'{u}nior} 
	 

%



\maketitle

\begin{abstract}
The increasing access to non-expensive sensors,  computing power, and more accurate forecasting of storm events provides unique opportunities to shift flood management practices from static approaches to an optimization-based real-time control (RTC) of urban drainage systems. Recent studies have addressed a plethora of strategies for flood control in stormwater reservoirs; however, advanced control theoretic techniques are not yet fully investigated and applied to these systems. In addition, there is an absence of a coupled integrated control model for systems composed of watersheds, reservoirs, and channels for flood mitigation.  

To this end, we develop a novel state-space model of hydrologic and hydrodynamic processes in reservoirs and one-dimensional channels. The model is tested under different types of reservoir control strategies based on real-time measurements (reactive control), and based on predictions of the future behavior of the system (predictive control) using rainfall forecasting. We apply the modeling approach in a system composed by a single watershed, reservoir, and a channel connected in series, respectively, for the San Antonio observed rainfall data. Results indicate that for flood mitigation, the predictive control strategy outperforms the reactive controls not only when applied for synthetic design storm events, but also for a continuous simulation. Moreover, the predictive control strategy requires smaller valve operations, while still guaranteeing efficient hydrological performance. From the results, we recommend the use of the model predictive control strategy to control stormwater systems due to the ability to handle different objective functions, which can be altered according to rainfall forecasting and shift the reservoir operation from flood-based control to strategies focused on increasing detention times, depending on the forecasting. 

\end{abstract}


\IEEEpeerreviewmaketitle
\newpage 
\section{Introduction}

\IEEEPARstart{F}{loods} are the deadliest natural disaster in US and worldwide \cite{wing2020new}. Estimated global flood damages from 1980 to 2019 exceed USD $750$ billion, with a peak in 2012 of nearly USD $70$ billion \cite{OurWorldindata2021}. Storm events are expected to become more frequent and intense due to climate change, likely increasing not only economic, but also social and environmental impacts, posing flood control as one of the greatest challenges for future planning and management of water resources systems \cite{gasper2011social}. Flood control measures in urban stormwater infrastructures are typically performed by static operations of valves, gates, pumps and/or tunnels based on pre-defined heuristic rules. With the advances in real-time control strategies such as the advent of non-expensive sensors, wireless communication, microprocessors and microcontrollers, opportunities to enhance flood management are evident. 
Therefore, control theory methods can be applied to enhance water resources management by deploying optimization-based control algorithms. Despite the fact that control theoretic methods have been applied to control combined sewer systems, reservoirs and drinking water systems \cite{Duchesne2001Mathematical,Troutman2020Balancing,Wang2020How}, it is a relatively new technique for drainage systems with separated infrastructure for stormwater and sanitary sewage \cite{Wong2018,Lund2018}.

Most drainage infrastructure in major cities was built to operate as static systems. These systems, in order to convey and/or store large storms (e.g., 100-yr storms), typically require relatively large dimensions. However, over the lifespan, the aging infrastructure, lack of proper maintenance, or the increasing of expected surface runoff (e.g., climate change and urbanization) can decrease the system reliability and hence increase the risk of flooding \cite{kessler2011stormwater,zhang2018urbanization}.  Real-Time Control (RTC) of the existent systems (e.g., watersheds, reservoirs, channels) can change the flow-storage regime by controlling actuators such as valves and pumps and ultimately restore or increase the level of protection against flooding. On the other hand, new stormwater systems designed for RTCs could require smaller surface areas and volumes, potentially leading to an overall cost reduction for the same level of expected performance \cite{Wong2018,brasil2021nature,xu2021enhancing}. 

\subsection{Literature Review}

The literature reviewed shows several applications of optimization of flow characteristics in a hydraulic structure (i.e., reservoir, channel, pipe, tunnel) in order to provide multiple benefits and increase the average performance to different water-related problems. An optimized control of the drainage facilities can enhance erosion control \cite{Schmitt2020}, provide stormwater runoff treatment due to higher detention times \cite{Sharior2019}, increase navigability conditions in canals \cite{horvath2014mpc}, and not only reduce flood downstream locally but also reshape the hydrographs in a desired way set by the developed optimization problem \cite{Wong2018}.

Another example of RTC approaches applied to water systems is on the topic of combined sewer overflows, which contains an extensive literature  \cite{Garcia2015,ocampo2010piece,joseph2015output}. However, only a few cities worldwide had applied optimization-based RTC in their drainage systems \cite{Lund2018}. Nonetheless, only a few studies assessed the benefits of RTCs in separated drainage systems. In general, there are 2 types of control strategies: Reactive Controls (i.e., based on Real-Time measurements), and Predictive Controls (i.e., based on predictions of the future behavior of the system). Reactive Controls can be used with heuristic or ruled-based approaches and with optimization-based approaches. Predictive Controls are typically solved with optimization-based approaches. Recently, \cite{Schmitt2020} assessed the role of heuristic and reactive control rules (i.e., based on Real-Time measurements)  applied to valves in stormwater reservoirs (i.e., actions are made based on measured states), focusing on control of erosive flows. The RTC efficiency was assessed through flow-duration curves, providing exceedance probabilities for any given flow. Although the RTC application provided a significant flow reduction for relatively small flows, its implementation for larger storms (i.e., $ > $ 1yr) had nearly less or equal peak flow reduction as the passive control. In several assessed storms, the outflows were larger than the inflows, indicating that the RTC increased the likelihood of large flows. Since the controls were guided to increase water quality by increasing detention times (e.g., the control algorithm principle relies on storing the water for one or two days for any particular event), when sequential storm events occur, flood risks increased because the storage capacity was nearly complete from previous storms.

The research conducted by \cite{Wong2018} presents an optimization methodology towards the control of stormwater reservoirs to identify optimal, but yet reactive control based on measured states. A Stormwater Management Model (SWMM) \cite{Rossman2010} model was used to estimate the flow dynamics in a watershed with several reservoirs and links. The model outputs were used in a linear quadratic regulator (LQR) control to decide the valve openings schedule in a set of controlled assets. Although the control schedule given by the LQR reduced peak flows in several ponds, a few of them had higher outflow peaks in some of the most intense storms assessed, mostly due to the lack of predictability of the future states of the system. Their results illustrate how RTC can increase flood control performance of urban drainage systems, but also indicate that without predictions, its practicality might be limited.


In contrast to heuristic and reactive controls, \cite{Shishegar2019} develop an optimization-based approach using a model predictive controller for operating valves in a stormwater detention pond. The authors used a calibrated SWMM model for the watershed and formulated the water balance dynamical problem in the reservoir by linear programming with flows as decision variables. Their approach reshaped hydrographs out of the stormwater reservoir according to the control objective, although several simplifications were adopted: (a) perfect 48-hours rainfall forecasting was used in the prediction horizon, (b) evaporation was neglected, and (c) linear hydraulics releasing outflows in a linear fashion. 

Other recent applications of RTC approaches using heuristic controls for water quality enhancement as the on/off and detention control \cite{Sharior2019}, fuzzy-logic and data-driven algorithms with genetic algorithms \cite{li2020data}, and deep learning \cite{mullapudi2020deep} are found in the literature and addresses different applications of RTC of urban drainage than our study. From these studies, we categorize RTC for separated urban drainage systems into (a) Static and/or Optimization-Based Reactive Controls (i.e., control algorithm according to measured or estimated states) and (b) Predictive / Optimization-Based Controls (i.e., control algorithms that considers future states estimation using rainfall forecasting and hydrological models). It is important to note that several ruled-based control (RBC) algorithms do not even require a hydrological modeling, although it is mandatory for predictive controllers \cite{Lund2018}.
\vspace{0.1mm}
\subsection{Paper Objectives and Contributions}
\begin{figure*}
    \centering
    \includegraphics[scale = 0.65]{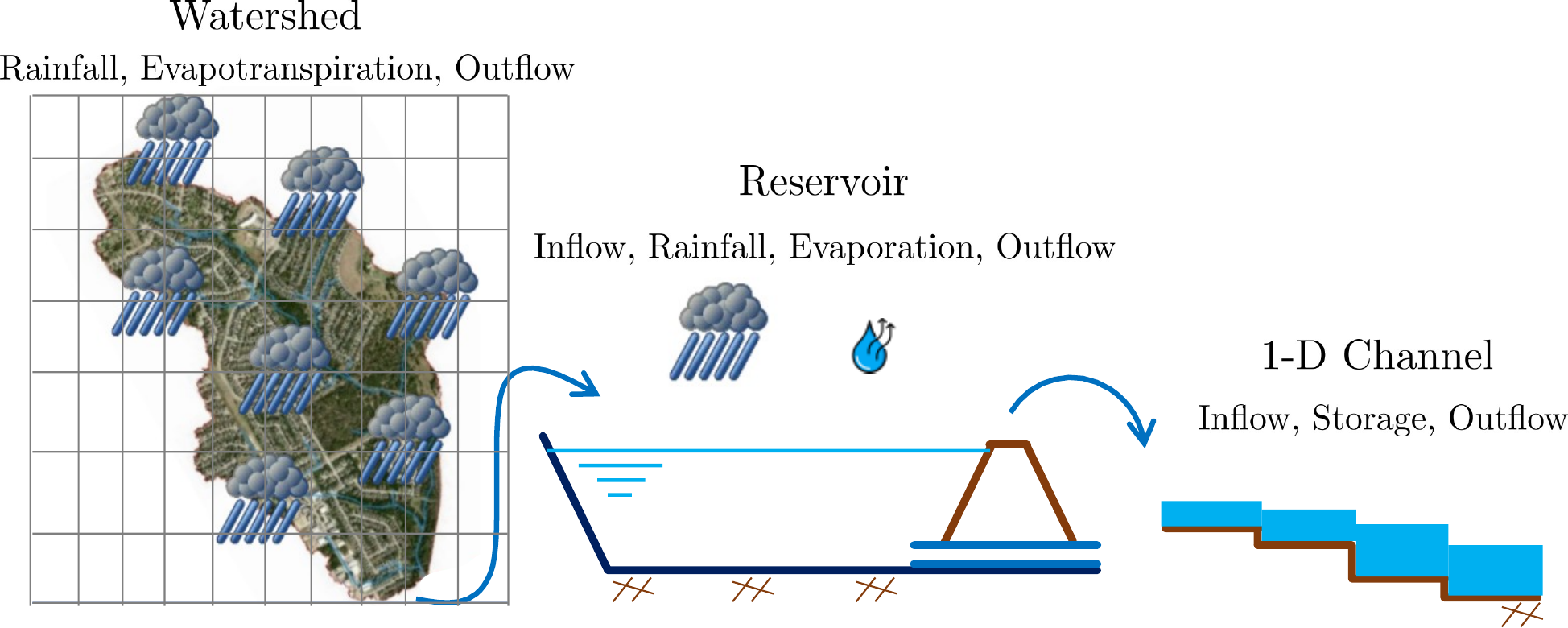}
    \caption{System of sub-systems where the Watershed is discretized in cells in a 2-D space, and the channel is discretized in 1-D cells}
    \label{fig:System}
\end{figure*}

We observe from the aforementioned studies a lack of coupled model for the main processes related to floods as overland flow, infiltration, reservoir routing, and channel routing that allows the use of advanced control theory techniques. Although SWMM or the Gridded Surface Subsurface Soil Analysis (GSSHA) are capable of solving the usually complex shallow water equations, their use in optimization for large systems can be intractable for real-time control and a simplified plant model of the flow dynamical model is needed \cite{Lund2018}. Moreover, exploring the differences between water quality controls as detention control, on/off control, further discussed in the next sections, into flood control performance are not yet investigated in the literature. In this paper, we address these issues and explore the trade-offs between reactive and predictive control strategies in regards to flood mitigation. A schematic of the modeled system is shown in Fig.~\ref{fig:System} representing the three modeled systems: watersheds discretized in cells, reservoirs receiving outflows from watershed, and channels discretized into sub-reaches receiving outflows from reservoirs. In this model, we are only interested in floods generated by excess of overland flow. Therefore, sub-surface flow are not considered and stored volumes are assumed as overland flow volumes.

To this end, we develop a novel state-space representation of the main processes of the water cycle related to urban catchments (i.e., infiltration, overland flow, reservoir storage, channel routing), using cells,  reservoirs, sub-reaches of channels connected as networked dynamical systems. This state-space model is based on energy, continuity, and momentum equations. We approach the non-linearities of the flow dynamics by performing successive linearizations in each time-step of the model using data from previous time-steps as operational points. Although we are able to linearize most of the system's equations based on physics laws (e.g., water balances, energy conservations), we are unable to do it for the rainfall intensity due to the complexity and absence of differentiable models for that. Therefore, we assume a known rainfall input time-series in the model and we model the watershed as a non-linear dynamical system.

In addition to the development of the novel state-space model, we want to assess how varied valve control strategies behave in the system. To that end, we develop a model predictive controller (MPC) algorithm to enhance the operation of valves in stormwater reservoirs, minimizing a composed cost function related to flood performance. Moreover, we compare the efficiency of reactive controls (i.e., controls based on real-time measurements of the states) with MPC solved with a gradient-based method (Interior-Point).

The fundamental contributions of this paper are described below:
\begin{itemize}
    \item{We present an overall mathematical representation including the flow dynamics in watersheds using the non-linear reservoir, and the Green-Ampt infiltration model \cite{Green1911Studies}, coupled with reservoir routing and 1-D channel dynamics in a non-linear state-space representation. It allows optimization and real-time control of urban drainage systems without requiring extensive software packages (i.e., only MATLAB is required)}.
    \item{We derive the non-linear dynamics of watersheds, and  linearize the dynamics of reservoirs and channels. Since no continuous and differentiable model is currently available for rainfall intensity, we assume a known time series of rainfall as a piecewise continuous input data.}
    \item{We provide a comprehensive analysis of reactive controls (i.e., some ruled-based and other optimally controlled) for flood mitigation and tested the efficiency of water quality ruled-based algorithms presented in \cite{Sharior2019} for flood mitigation}. 
    \item{We develop and apply a servo-control algorithm \cite{young1972approach} used in the discrete linear quadratic integrator (DLQI) reactive control. This is a new application for this control technology in urban drainage systems. This algorithm allows tracking a specific state and can be used in reservoirs with minimum specified water surface depths (e.g., wetlands or retention ponds)}.
    \item{We evaluate and discuss the performance of reactive RBCs (passive, on/off, detention control) and reactive optimization-based controls (discrete linear quadratic regulator and discrete linear quadratic integrator) compared with the predictive control strategy. Moreover, we explore the caveats of consecutive design storms and also compared the flood performance of reactive and predictive algorithms in a continuous simulation, providing a methodology to assess the efficiency of control algorithms for flood mitigation. }
\end{itemize}

The remainder of the paper is organized as follows. Section~\ref{sec:Materials} develops the state-space model for watersheds, reservoirs, and 1-D channels flow routing dynamics. Next, Section~\ref{sec:static_controls} describes the tradional reactive controls applied to control of stormwater reservoirs. Moreover, in this section we develop a novel non-linear MPC optmization problem focusing on flood mitigation in channels and reservoirs. Following, in Section~\ref{sec:Model_Application} we present a case study to test the controls presented in the aforementioned sections, including two scenarios: 2 consecutive design storms of 25-yr, 12-hr and 10-yr, 12-hr, respectively and a continuous simulation scenario from 04/23/2021 to 07/23/2021 in San Antonio - Texas. Section~\ref{sec:Results} shows the results and discussion of the model application and Section~\ref{sec:Conclusions} the conclusions, limitations, and future works. The paper notation for this paper is introduced next.

\vspace{-0.0cm}
\noindent \textit{\textbf{Paper's Notation:}} 
Italicized, boldface upper and lower case characters represent matrices and column vectors: $a$ is a scalar, $\m a$ is a vector and $\m A$ is a matrix. Matrix $\m I_n$ denotes an identity square matrix of dimension $n$-by-$n$, whereas $\m O_{m \times n}$ and $\m 1_{m \times n}$ denotes a zero and one matrix with size $m$-by-$n$, respectively.
The notations $\mathbb{R}$ and $\mathbb{R}_{++}$ denote the set of  real and positive real numbers. Similarly, $\mathbb{N}$ and $\mathbb{N}_{++}$ denote the set of natural and positive natural numbers. The notations $\mathbb{R}^n$ and $\mathbb{R}^{m\times n}$ denote a column vector with $n$ elements and an $m$-by-$n$ matrix in $\mathbb{R}$. The element-wise product or Hadamard product is defined as $\m x \circ \m y \coloneqq [x_1 y_1, x_2 y_2, \dots{}, x_n y_n]^T$ multiplications. Similarly, the element-wise division or Hadamard is defined as $\m x \oslash \m y \coloneqq [ \frac{x_1}{y_1},\frac{x_2}{y_n},\dots{},\frac{x_n}{y_n}]^T $.
The element-wise $p$ power of a matrix $\m A, $ $(\m A^{\circ p})$, with $\m A \in \mathbb{R}^{m \times n}$ and $p \in \mathbb{R}$  is given by $ a_{i,j}^p$ for $i \in \mathbb{N}_{++}$, and $ j \in \mathbb{N}_{++}$
The number of elements in a set $\m A \cup \m B$ is $\textbf{n}(\m A \cup \m B) = \textbf{n}(\m A) + \textbf{n}(\m B) - \textbf{n}(\m A \cup \m B)$. A normally distributed random number with average $\mu$ and variance $\m \sigma^2$ is notated by $\mathcal{\m N}(\mu,\sigma^2)$. Given a vector $\m x \in \mathbb{R}^n$, the notation $\m x(i:j)$ with $i$ and $j \in {N}_{++}$ represents a cut in $\m x$ from i\textsuperscript{th} to j\textsuperscript{th} entries.

\section{Mathematical Model Development} ~\label{sec:Materials}
The stormwater flow dynamical problem solves physics-based governing equations in each watershed, reservoir, and channel using physically-based input data. Since mass balance equations are solved, we postulate the stormwater flow dynamical system as nonlinear difference-algebraic equation (DAE) state-space model. All variables used in this paper are summarized in Tab.~S1 in the supplemental material. 


The model requires matrices and vectors to represent the dynamical and algebraic parts of the simulated hydrological systems, as presented in Eq.~\ref{equ:statespace}. Specifically, matrix $\m E$ enables the representation of the dynamics and algebraic constraints of the coupled watershed-reservoir-channel model in a single state-space model. Moreover, linear time-varying parts are represented by matrices $\m A(k), \m B(k),$ while $\m C$ represents time-invariant output matrix, as the sensors are assumed to have fixed geographic placements. In addition, offsets, non-linearities from the watershed model, operational points, rainfall intensity in each cell, outflows/inflows connectivity from watersheds to reservoirs and from reservoirs to channels, and integrator reference setpoints are given by $\m \psi (\cdot)$.  The mathematical development of these matrices and vectors are detailed in the following sections. In this paper, we develop the DAE state space model for a system composed of a single watershed, reservoir, and channel. We collect a vector of water surface depths in cells, reservoirs, and channels, accumulated infiltration depths in each cell, and outflows from catchments and reservoirs as the state vector, such that $\m x(k) = [\m h_{ef}^w(k),\m f_d^w(k), q_{out}^w(k), h^r(k), q_{out}^r(k), \m h^c(k)]^T$. We also assume a control vector given by $\m u(k) = u^r(k)$. The model parameters, states, outputs, and sources of uncertainty are presented in Table~\ref{tab:model_parameters}. The state-space representation can be written as 
\begin{subequations} ~ \label{equ:statespace}
\begin{align} 
  \m E \m x(k+1)& =\m A(k)\m x(k) + \m B(k) \m u(k) + \m \psi(k,\m x(k), \m x^r_{*},\m u^r_{*}) \\
  \m y(k) &= \m Cx(k)
\end{align}
\end{subequations}
where $\m E$ is a singular matrix with some zero rows representing the algebraic constraints of flow equations, $\m A(k) \in \mbb{R}^{ n \times n}$, is the state or system matrix, $\m x(k) \in \mbb{R}^n$ is the state vector, $\m B(k) \in \mbb{R}^{n \times m}$ is the input matrix, $\m u(k) \in \mbb{R}^m$ is  the input vector, $\m \Phi(k,\m x^r_{*},\m u^r_{*})$ is a disturbance vector, $\m x^r_{*}$ and $\m u^r_{*}$ are operational points, $ \m C \in \mbb{R}^{p \times m}$ is the output matrix, and $\m y(k)$ is the output function and k is a time-step index. The vectors $\m h_{ef}^w, \m h^r,$ and $\m h^c$ are water surface depths in each system, $\m F_d$ is the accumulated infiltration depth in the cells of the watershed, and $Q_{out}^w$ and $Q_{out}^r$ is the watershed outflow, and reservoir outflow, respectively. In the following sections, we define each system and their governing equations, as well the linerizations. 

\begin{table*}
\small
\centering
\caption{Systems, Parameters, States, Outputs and Uncertainty features of the model, where $q$, $n_r$ and $n_c$ represents the number of cells, reservoirs and channel sub-reaches, and the state vector dimension $n$ is equal $(2q + 1 + 2n_r + n_c$).}
\label{tab:model_parameters}
\begin{tabular}{lllll} 
\hline
System                      & Parameters                                                               & States  (Symbols)                                                                                                                                                                                              & Outputs  (Symbols)                                                                                                                                                            & Uncertainty                                                                                                          \\ 
\hline\hline
\multirow{5}{*}{Cells}      & Infiltration Parameters                                                  & \multirow{5}{*}{\begin{tabular}[c]{@{}l@{}}Water Surface Depth in each\\ cell ($\m h_{ef} \in \mbb R^{q}$)\\in (mm), Infiltrated Depths \\($\m f_d \in \mbb{R^q}$) in (mm),\\and Outflow $(q_{out}^w)$ in (m\textsuperscript{3}/s)\\\end{tabular}} & \multirow{5}{*}{\begin{tabular}[c]{@{}l@{}}Outflow in each\\cell and in the outlet \\($\m q_{out}\in \mbb R^{q}$)\\in (mm.h\textsuperscript{-1})\end{tabular}}                & \multirow{5}{*}{\begin{tabular}[c]{@{}l@{}}Manning's\\Coefficient and\\Rainfall Spatial\\Distribution\end{tabular}}  \\
                            & Surface Roughness                                                        &                                                                                                                                                                                                                &                                                                                                                                                                               &                                                                                                                      \\
                            & Initial Abstraction                                                      &                                                                                                                                                                                                                &                                                                                                                                                                               &                                                                                                                      \\
                            & DEM                                                                      &                                                                                                                                                                                                                &                                                                                                                                                                               &                                                                                                                      \\
                            & LULC                                                                     &                                                                                                                                                                                                                &                                                                                                                                                                               &                                                                                                                      \\ 
\hline
\multirow{2}{*}{Reservoirs} & \begin{tabular}[c]{@{}l@{}}Stage-Discharge \\ Relationships\end{tabular} & \multirow{2}{*}{\begin{tabular}[c]{@{}l@{}}Water Surface Depth\\($\m h^r \in \mbb R^{n_r}$) in (m),\\and Outflow $q_{out}^r$ in (m\textsuperscript{3}/s)\end{tabular}}                                                                  & \multirow{2}{*}{\begin{tabular}[c]{@{}l@{}}Maximum Water Surface Depth \\($\max(\m h^r)\in \mbb R$) in (m)\end{tabular}} & \multirow{2}{*}{\begin{tabular}[c]{@{}l@{}}Water Surface Depth \\Measurement\\Noise\end{tabular}}                            \\
                            & \begin{tabular}[c]{@{}l@{}}Area-Volume Function \\Porosity\end{tabular}  &                                                                                                                                                                                                                &                                                                                                                                                                               &                                                                                                                      \\ 
\hline
\multirow{4}{*}{Channels}   & \begin{tabular}[c]{@{}l@{}}Stage-Discharge \\ Relationships\end{tabular} & \multirow{4}{*}{\begin{tabular}[c]{@{}l@{}}Water Surface Depth \\($\m h^c \in \mbb R^{n_c}$) in each\\sub-reach of the\\Channel (m)\end{tabular}}                                                              & \multirow{4}{*}{\begin{tabular}[c]{@{}l@{}}Maximum Water Surface Depth\\($\max(\m h^c)\in \mbb R^{n_c}$)\\in (m)\\\end{tabular}}                                                      & \multirow{4}{*}{\begin{tabular}[c]{@{}l@{}}Manning's\\Coefficient\end{tabular}}                                      \\
                            & Manning's Coefficient                                                    &                                                                                                                                                                                                                &                                                                                                                                                                               &                                                                                                                      \\
                            & Hydraulic Radius                                                         &                                                                                                                                                                                                                &                                                                                                                                                                               &                                                                                                                      \\
                            & Gridded Bathymetry                                                       &                                                                                                                                                                                                                &                                                                                                                                                                               &                                                                                                                      \\
\hline\hline
\end{tabular}
\end{table*}

For cases with more watersheds, reservoirs, and channels, the state vector can be augmented to include the new systems, concatenating each individual state (e.g., $\m h_{ef}^w(k) = [\m h_{ef,1}^w, \dots, \m h_{ef,s}^w]^T $) such that $\m x(k) = [\m h_{ef}^w(k),\m f_d^w(k), \m q_{out}^w(k), \m h^r(k),\m q_{out}^r(k)]^T$, $\m u(k) = [u_1^r(k), \dots u_s^r(k)]^T$, where $s$ is the number of systems composed by watersheds, reservoirs, and channels. In this paper, we present the mathematical formulation for a watershed-reservoir-channel system.
\subsection{Watershed Overland Flow Modeling} ~\label{sec:overland_flow}
In this section we derive the mathematical formulation to estimate overland flow in watersheds using a fully-distributed hydrologic model. Excess of infiltration (Hortonian Flow) and/or saturation (Dunnian Flow) generates overland flow in storage cells \cite{Maxwell2014}. In urban environments with high impervious areas, hortonian flows governs the overland flow generation and occur when infiltration capacity is smaller than the inflow rate (i.e., net precipitation, inflow from neighbour cells, ponding depth).

The infiltration losses are estimated using the Green and Ampt Infiltration Model. This model is physically-based and derived from simplifications of the Richards' Equation \cite{Richards1931Capillary,Green1911Studies}. All parameters of the model can be estimated in laboratory tests. However, substantial studies are available in the literature providing good parameter's estimates according to the soil characterization  \cite{Green1911Studies,Rossman2010}. Furthermore, Green-Ampt soil properties are typically available in spatially distributed Geographical Information System (GIS) databases in many parts of the world, which facilitates the application of this model. 

For each cell of a pre-defined grid domain, the (i) saturated hydraulic conductivity, (ii) suction head pressure (capillarity), (iii) initial and saturated soil moisture and (iv) initial infiltrated water content are required. Extractions of ASCII files from delineated watershed rasters can be used to define the matrices representing the digital elevation model (DEM) and the imperiousness map. More details of these files and for the model construction are found in \cite{GomesJr.2021}. The Manning’s equation is used to relate water surface depth to flow for the cells \cite{Akan1993Urban,Chow2010Applied}. To perform the calculations, input data such as (a) Manning’s coefficient and (b) Central Elevation of each cell of the grid are required. 
\begin{figure}[t]
\vspace{-0.4 cm}
    \centering
    \includegraphics[scale = 0.55]{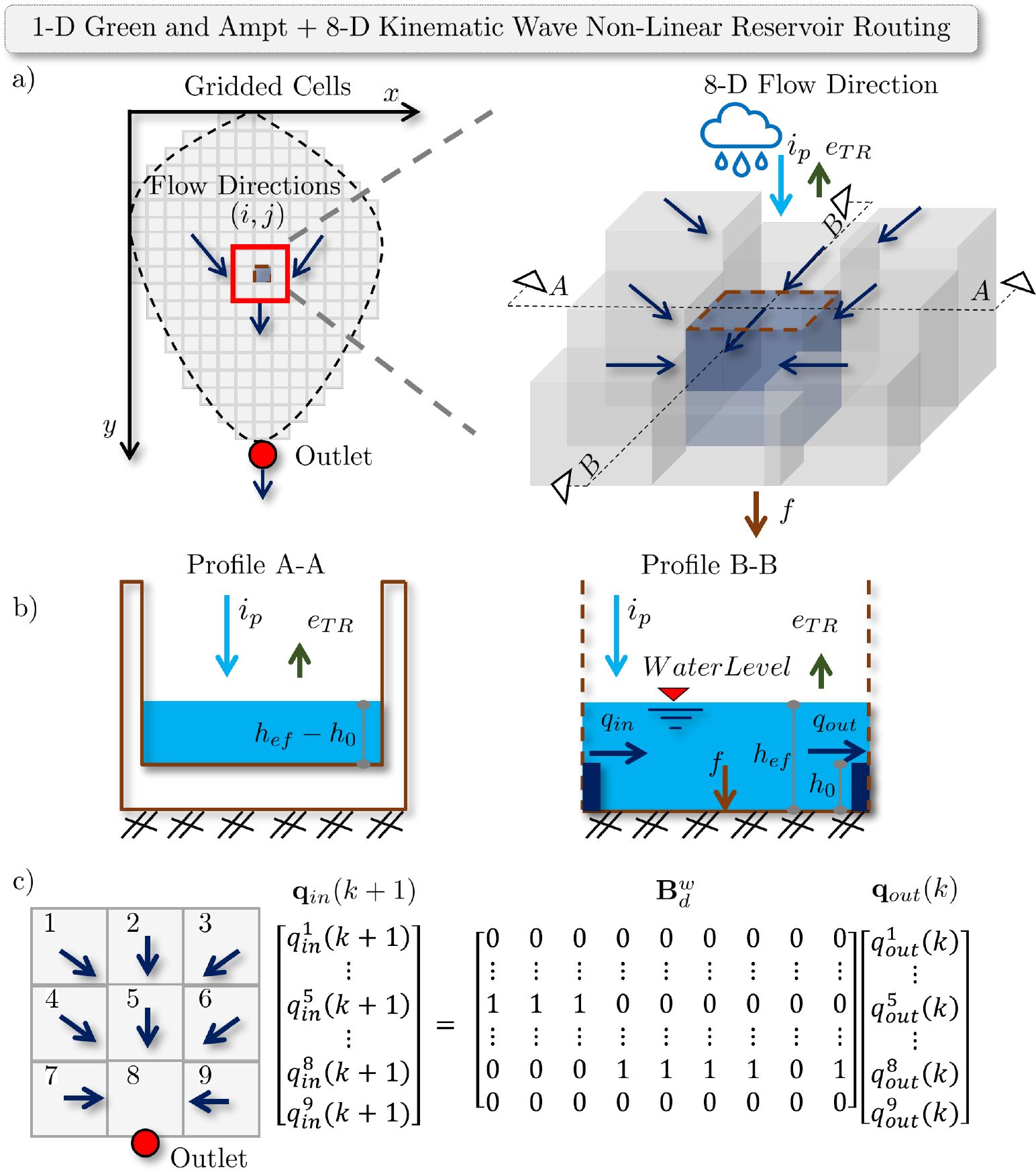}
    \caption{(a) Watershed conceptual model, $a)$ is the plan view of the watershed, $b)$ is the profiles of a cell and $c)$ is the $8$-$D$ flow direction matrix based on the steepest slope, where $q_{in}, I, ETR, q_{out}, f, d $ and $ h_0$ are the inflow, rainfall intensity, evapotranspiration, outflow, infiltration, water surface depth and the initial abstraction of a specific cell of the grid.}
    \label{fig:wshedmodel}
\end{figure}
\subsubsection{1-D Vertical Infiltration Model}
The Green-Ampt model is applied to each grid cell to estimate the infiltration capacity at any given time and is used to estimate the available depth to be routed to downstream cells. The infiltration capacity is estimated as
\begin{equation} \label{equ:green_ampt}
    c^{i,j}(t) = k_{sat}^{i,j}\left[1 + \frac{\left(\zeta^{i,j} + h_{ef}^{i,j}(t)\right)\left(\theta^{i,j}_s - \theta^{i,j}_i\right)}{f_d^{i,j}(t)}\right],
\end{equation}
where the sub-index $i$ and $j$ indicate the cell position in the grid, $c(t)$ is the infiltration capacity (mm.h\textsuperscript{-1}), $k_{sat}$ is the saturated hydraulic conductivity in (mm.h\textsuperscript{-1}), $\zeta$ is the suction head pressure (mm), $h_{ef}(t)$ is the water depth in the cell (mm), ($ \Delta \theta = \theta_s - \theta_i$) is the effective soil moisture, $f_d(t)$ is the time-varying accumulated infiltration (mm).

The infiltration model is a non-linear time-varying function of the accumulated infiltrated volume $I_{i}^{i,j}(t)$, and is dynamically computed in an explicit 1\textsuperscript{st} order finite difference discretization, written as
\begin{equation} \label{equ:infiltration}
\begin{split}
    f_d^{i,j}(t+\Delta t) &= f_d^{i,j}(t) + f^{i,j}(t)\Delta t \\
    = f_d^{i,j}(t) +   &\overbrace{\left[\min{\left(c^{i,j}(t),q_{in}^{i,j}(t) + i_p^{i,j}(t) - e_{TR}^{i,j}(t) \right)}\right] }^{f^{i,j}(t)}\Delta t,
\end{split}
\end{equation}

\noindent{where $i_p$ is the rainfall intensity (mm.h\textsuperscript{-1}), $e_{TR}$ is the real evapotranspiration intensity (mm.h\textsuperscript{-1}), $\Delta t$ is the model time-step, $f_d^{i,j}(t)$ is the accumulated infiltration depth in (mm), $q_{in}^{i,j}(t)$ is the inflow discharge rate (mm.h\textsuperscript{-1}), and $h_{ef}^{i,j}$ is the runoff water depth (mm) in cell $i,j$}. 

In the previous equation, we model the soil drying only by assuming a flux of evapotranspiration occurring at pervious surfaces such that in drying periods, the soil storage depth is decreased. Moreover, we limit the accumulated infiltrated soil depth $f_d$ to a minimum threshold typically assumed as 5 mm.

\subsubsection{Vertical Water Balance}
Hydrological processes of infiltration, precipitation, surface runoff and evaporation occur simultaneously. However, an analytical solution of the continuous functions of these inputs into the overall water balance is typically not available for real case scenarios, especially due to the rainfall. With proper stable time-step resolution, an alternative to solve the overland flow dynamics is to derive an explicit system of equations from the time derivative of the storage variation in each cell, expressed as follows \cite{Rossman2010}

\begin{subequations}
\footnotesize
\begin{align}
\dv{s^{i,j}(t)}{t} &= \Bigl[ q_{in}^{i,j}(t) + i_p^{i,j}(t) - e_{TR}^{i,j}(t) - q_{out}^{i,j}(t) - f^{i,j}(t) \Bigl] 
 \omega^w  \\
 \dv{h_{ef}^{i,j}(t)}{t} &= \frac{1}{\omega^w}\dv{s^{i,j}(t)}{t} \label{equ:diff_storage},
\end{align}
\end{subequations}

\noindent where $s^{i,j}$ is the ponded water storage (m\textsuperscript{3}), $\omega^w = \Delta x^w \Delta y^w$ is the cell area in (m\textsuperscript{2}), $\Delta x^w$ and $\Delta y^w$ are the cell resolution in $x$ and $y$ in (m), $q_{out}^{i,j}(t)$ is the outflow discharge rate (mm/hr) and $f^{i,j}(t)$ is the infiltration rate (mm/hr).


Generalizing Eq.~\eqref{equ:diff_storage} for a vector notation concatenating the number of rows and columns into a vector of dimension $q$, changing the units for water surface depth instead of storage, and including a constraint to account for the initial abstraction, we can derive the following expression 
$$ \m h_{\m ef}(k+1) = \m h_{\m ef}(k) $$
\vspace{-0.5cm}
\begin{equation} \label{equ:finite_diff}
   + \frac{\Delta t}{600}\Bigl(\m q_{in}(k) + \m i_p(k) - \m e_{TR}(k)  -  \m q_{\m out}(k) - \m f(k)\Bigr), 
\end{equation}
\noindent{where $k$ is a time-step index, $\m h_{ef}$, $\m q_{in}$, $\m i_p$, $\m e_{TR}$, $\m q_{out}$, and $\m f$ $ \in \mbb{\m R}^{q}$.}

\subsubsection{Outflow Discharge}
The outflow discharge is simplified to be a function of the steepest 8-Direction slope, cell roughness and water surface depth in a kinematic-wave shallow-water simplification approach. This approximation is implemented since the goal of the watershed model is to determine flows and not high-resolution surface water flood depths. According to the gridded elevation, first we define a flow direction matrix similar as presented in Fig.~\ref{fig:wshedmodel}, part b). This matrix is determined calculating the steepest topographic slope of each cell assuming 8 boundary cells (i.e., Moore neighborhood grid). According to the steepest slope direction, a number is assigned to each cell. This matrix defines the boundary conditions among each cell in the grid. 

Using the Manning’s equation \cite{Chow2010Applied}  and assuming the energy slope as bottom slope, we can estimate the outflow discharge for a specific cell as following in a matrix notation
\vspace{-0.2 cm}
\begin{equation} \label{equ:manning}
\footnotesize
     \m q_{out}(k)  = \overbrace{\Bigl(k_f \frac{\Delta x^w + \Delta y^w}{2}\Bigr){\m s_{0}^{\circ 1/2}} \oslash \m n}^{\m \lambda} \circ  \Bigl(\max(\m h_{ef}(k) - \m h_0),0 \Bigr)^{\circ {5/3}}
\end{equation}
where $k_f$ is a conversion factor equals $1\times 10^{-5}$,   $\m s_{\m {0}}$ is the steepest topographic slope (m/m), $n$ is the Manning’s roughness coefficient in (sm\textsuperscript{-1/3}), and $\m \lambda$ lumps the hydraulic properties of the cells into a single vector. $\m s_{0}$, $\m h_0$, $\m q_{out}$, and $\m \lambda \in \mbb{R}^q$. All the other equations are in the international system of units.

To guarantee continuity in the hydrological model, a pre-processing in the digital elevation model is performed. Natural sinks are filled to ensure all cells have an outlet slope. Moreover, the outlet boundary condition of the watershed is modeled assuming a normal flow hydraulic condition \cite{Kollet2006}.
\subsubsection{Inflow Discharge}
The inflow discharge is a function of the flow direction matrix and outflow discharge. Defining a matrix $\mB_{d}^w \in \mbb{R}^{q \times q}$ to represent the direction relationship among the cells in the form of a sparce matrix filled with ones (i.e., outflows becomes inflow for the downstream cell)  and zeros (i.e., no flow connection between cells, see Fig.~\ref{fig:wshedmodel} part c), we can write
\vspace{-0.3 cm}
\begin{equation} \label{equ:disturbance}
    \m q_{\m in}(k) = \m B_d^w \m q_{\m out}(k),
\end{equation}
where $\mB_{\m q} \in \mathbb{R}^{q \times q}$ is the direction boolean matrix containing the relationship between each cell. 

Substituting Eqs.~\eqref{equ:disturbance} and \eqref{equ:manning} into Eq.~\eqref{equ:finite_diff}, tracking the accumulated infiltration depth ($ \m f_d$), and the watershed outflow ($q_{out}^w$), the watershed sub-system from Eq.~\eqref{equ:statespace} can be written in a state-space representation, given by
\begin{strip}
\hrule
\begin{equation} \label{equ:state_watershed}
\begin{split}
\overbrace{
\begin{bmatrix}
    \m I_{2q} & 0 \\
     0 & 0
\end{bmatrix}
}^{\m E^w}
\overbrace{
\begin{bmatrix}
    \m h_{ef}(k+1) \\
    \m f_d(k+1) \\
    q_{out}^w(k+1)
\end{bmatrix}
}^{\m x^w(k+1)}
= \overbrace{\m I_{2q+1}}^{\m A^w(k)}\m x^w(k) + \overbrace{\Delta t 
    \begin{bmatrix}
      1/600\Bigl((\m B_d^w - \m I_q)\circ \m \lambda \circ \max\Bigl(\m h_{ef}(k) - \m h_0,0\Bigr)^{\circ 5/3} + \m i'_p(k) - \m f(k) \Bigr) \\
      \m f(k) \\
      \frac{-0.277 \times 10^{-6}}{\Delta t}|| \m \lambda^{\m i_0, \m j_0} \max\Bigl(\m h_{ef}^{\m i_0, \m j_0}(k) - \m h_0,0\Bigr)^{\circ 5/3} \omega^w ||
    \end{bmatrix}
    }^{\m \psi^w(k,\m x^w(k))},
\end{split}
\end{equation}
\hrule
\vspace{-0.5cm}
\end{strip}

\noindent where $\m x^w(k) = [\m h_{ef}^w(k), \m f_d^w(k), q_{out}^w(k)]^T$, $\m i'(k) = \m i_p(k) - \m e_{TR}(k)$, $\m f(k)$ is modeled in Eq.~\eqref{equ:infiltration}, $\m i_0$ and $\m j_0$ represent the indexes of the outlet cells, and $\m \lambda^{i_0,j_0}$ concatenates $\lambda$ for outlet cells.
\subsection{Reservoir Dynamics} ~\label{sec:reservoir_dynamics}
In this section, the reservoir routing dynamics is described and we provide a fully linearizable model to account for valve control in stormwater reservoirs with hydraulic devices as orifices and spillways.
\subsubsection{Orifice Modeling}
The control signal $u(k)$ represents the percentage of the orifice area that allows flow to be routed to downstream channels. Therefore, applying the energy equation in the reservoir \cite{Chow2010Applied} and including $u(k)$ in the effective orifice area, we can derive the controlled orifice equation, such that 
\begin{equation}\label{equ:orifice}
\footnotesize
\begin{split}
q_o(h^r(k),u(k)) &= u^r(k)c_{d,o}a_{o}\sqrt{2g(\max(h^r(k)-  (h_{o} + h_{m}),0))} \\
&=  u^r(k)k_o\sqrt{\hat{h}^r(k)},
\end{split}
\end{equation}
where $q_o$ is the orifice discharge, $c_{d,o}$ is the orifice discharge coefficient, $a_o$ is the orifice area and $g$ is the gravity acceleration, $\hat{h}^r(h^r(k)) = \max(h^r(k)-  (h_{o}+h_{m}),0)$ is the effective water depth at the orifice, $u$ is the control input representing the valve opening between $0$ and $1$, $h_o$ is the bottom elevation of the orifice and $h_m$ is the minimum water surface depth to begin the outflow (i.e., typically 20\% of the hydraulic diameter of the outlet) \cite{Chow2010Applied}.
\subsubsection{Spillway Modeling}
The spillway is also assumed to be discharging at the atmospheric pressure. The Francis Spillway equation is typically used for detention reservoirs and can be modeled as follows
\begin{equation} \label{equ:spillway}
    q_{s}(h^r(k)) = c_{d,s}l_{ef}(h^r(k)-p)^{3/2} = k_s(h^r(k)-p)^{3/2},
\end{equation}
where $q_s$ is the spillway discharge, $c_{d,s}$ is the spillway discharge coefficient and $l_{ef}$ is the effective length of the spillway \cite{Chow2010Applied}.
\subsubsection{Reservoir Outflow}
The outflow in a reservoir is a function of the water surface depth $ h^r(k)$ and is described by the energy conservation applied into the orifice and spillway, and thus has two governing equations. The first case is where $h^r(k) \leq p$, and is given by Eq.~\eqref{equ:orifice} \cite{Chow2010Applied}. When the water level reaches the spillway level, the reservoir outflow ($q_{out}^r$) is the sum of the orifice and spillway flow. Therefore, the reservoir outflow function be derived as follows
\begin{equation}
    q_{out}^r\left(h^r(k),u^r(k)\right) = 
    \left\{\begin{array}{l}u^r(k)k_o\sqrt{\hat{h}^r(k)}\ \  \textbf{if} \ \
	h^r(k)\leq p,\ \ \textbf{else} \\
	u^r(k)k_o\sqrt{\hat{h}^r(k)}+k_s{\left(h^r(k)-p\right)}^{3/2}\end{array}\right.
\end{equation}

We compute the jacobian of $q_{out}^r$ with respect to $h^r$ and $u$ to obtain a linearized flow equation neglecting the high order terms of the Taylor's series, resulting in the following equations:

\begin{strip}
\vspace{-0.37cm}
\hrule
\begin{equation} \label{equ:diff_reservoir}
    \frac{\partial q_{out}^r(h^r(k),u^r(k))}{\partial h} \approx 
    \left\{ \frac{u^r(k)k_o}{2\sqrt{\hat{h}^r(k)}} + \frac{3k_s [\max((h^r(k)-p)^{1/2},0)]}{2}\right\} = 
    \alpha (h^r(k),u^r(k))
\end{equation}

\begin{equation} \label{equ:beta}
    \frac{\partial q_{out}^r(h^r(k),u^r(k))}{\partial u} = k_o\sqrt{\hat{h}^r(k)} = \beta(h^r(k),u^r(k)).
\end{equation}

Therefore, a linearized model for the outflow in terms of the stored water surface depth and valve opening is given as
\begin{equation} \label{equ:linearized1}
    q_{out}^r\left(h^r(k),u^r(k)\right)=\overbrace{q_{out}^r\left(h^r_{*}\right)}^{\gamma(k)}+\overbrace{\left.\alpha\right\vert_{h=h^o,u=u^r_{*}}}^{\tilde{\alpha}(k){}}\left(h^r(k)-h^r_{*}\right) + 
    \overbrace{{\left.\beta\right\vert{}}_{h=h^o,u=u^r_{*}}}^{\tilde{\beta}(k){}}\left(u^r(k)-u^r_{*}\right),
\end{equation}
\hrule
\vspace{-0.5cm}
\end{strip}
\noindent{where $\gamma$ is the offset, $\tilde{\alpha}$ is the linear coefficient with respect to $h^r$, $\tilde{\beta}$ is the linear coefficient in terms of $u$ and $u^r_{*}$ and $h^r_{*}$ are operation points given by the states and controls of the previous time-step, such that $u^r_{*}(k) = u^r(k-1)$ and $h^r_{*}(k) = h^r(k-1)$ .}
\subsubsection{Reservoir Water Balance}
The temporal evolution of storage in a reservoir depends on the inflow, precipitation, evaporation, water surface area and stage-discharge function of the outlet hydraulic devices. Applying evaporation and precipitation in the reservoir surface area, we can derive an expression for the water storage dynamics, given by 
\vspace{-0.3 cm}
\begin{align}\label{equ:reservoir_balance}
\small
\frac{\partial s^r(h^r(k),u^r(k))}{\partial t} &= \overbrace{q_{out}^w(k) + \Bigl(i(k) - e_v(k)\Bigr) \omega^r(h^r(k))}^{q_{in}^r(k,h^r(k))} \notag \\
&-~ q_{out}^r(h^r(k),u^r(k)), 
\end{align}
where $s^r$ is the stored volume of stormwater runoff in the reservoir, $\omega^r(h^r(k))$ is the reservoir surface area in terms of $h^r(k)$, $e_v$ is the evaporation in the reservoir surface area, $q_{out}^w$ is the inflow from upstream catchment, $q_{in}^r$ is the total inflow $i$ is the rainfall intensity.
Assuming an average porosity $\eta$ representing the stage-storage relationship in the reservoir (e.g., for free surface reservoirs, the porosity is 1), the water surface depth dynamics can be derived as
\begin{align} \label{equ:porosity}
\frac{\partial h(k,h^r(k),u^r(k))}{\partial t}  =& \frac{1}{\omega^r(h^r(k))\eta}\Bigl[q_{in}^r(h^r(k)) 
 ~- \\  &q_{out}^r(h^r(k),u^r(k))\Bigr].
\end{align}
The storage dynamics in a reservoir can be very slow depending on the area of the reservoir, which might contribute for low degrees of controllability for reactive controls, even in events with high inflows. Substituting the linearized reservoir outflow, Eq.~\eqref{equ:linearized1}, into the water balance equation, it follows that
\vspace{-0.4 cm}
\begin{align} \label{equ:h_dot}
\tiny
&\frac{\partial h(h^r(k),u^r(k))}{\partial t} = \overbrace{\frac{1}{{\omega^r(h^r(k))}\eta}}^{\mu(h^r(k))}\Bigl[q_{in}^r(k,h^r(k))  \notag \\ & -  \tilde{\alpha}(k)(h^r(k)-h^r_{*})  - \tilde{\beta}(k)(u^r(k) - u^r_{*}) - \gamma(k) \Bigr].
\end{align}
Assuming an approximated finite-difference scheme by the forward Euler method applied in the water surface depth partial derivative equation, we obtain
\begin{equation} \label{equ:euler}
    \frac{\partial h(h^r(k),u^r(k))}{\partial t} \approx \frac{h^r(k+1) - h^r(k)}{\Delta t}.
\end{equation}
Generalizing the reservoir dynamics for more than one reservoir per watershed ($n_r > 1$ and $\m u^r(k) \in \mathbb{R}^{n_r}$), substituting Eq.~\eqref{equ:h_dot} into Eq.~\eqref{equ:euler}, and expanding for a matrix notation, the water depth dynamics in reservoirs is given by Eq.~\eqref{equ:state_reservoir}, the reservoirs sub-system from Eq.~\eqref{equ:statespace} is given by
 
 \begin{strip}
 \hrule
\begin{align} \label{equ:state_reservoir}
    \m h^r\left(k+1\right) =& \overbrace{\left(\m I_{\m n_r}-\mr{diag} \left(\Delta t \m \tilde{ \m \alpha}(k) \circ  \m \mu(\m h^r(k))  \right)\right)}^{\tilde{\m A}^r(k)}\m h^r\left(k\right) +\overbrace{(-\mr{diag}(\Delta t \m \tilde{\m \beta}(k) \circ \m \mu(\m h^r(k))))}^{\tilde{\m B}^r(k)}\m u^r(k) + \\ &\overbrace{\Delta t \m \mu(\m h^r(k)) \circ \left(\m \tilde{\m \alpha{}}(k) \circ \m h^r_{*} + \tilde{\m \beta{}}(k) \circ \m u^r_{*}-\m \gamma(k){}+ \m q_{in}^r\left(k\right)\right)}^{{ \tilde{\m \psi}{}}^r(k,\m u^r_{*},\m x^r_{*})},
\end{align}   
 \hrule

\noindent where $\m h^r(k),\tilde{\m \alpha}(k),\tilde{\m \beta}(k)$ and $\tilde{\m \psi_r}(k,\m u^r_*, \m x^r_*) \in \mbb{R}^{n_r}$,  $n_r$ is the number of reservoirs, $\tilde{\m A}(k)$  and $\tilde{\m B}_r(k) \in \mbb{R}^{n_r \times n_r}$ and $\m q_{in}^r(k)$ captures the watershed-reservoir outflow/inflow connection.

Moreover, generalizing the method for $s$ systems with $n_r$ reservoirs per watershed, we can define $z = n_rs$ resulting in the reservoir state space non-linear dynamics tracking the reservoir outflows such that:
\vspace{-0.3cm}
{\small \begin{equation} \label{equ:state_space_reservoir}
\begin{split}
    &\overbrace{
    \begin{bmatrix}
        \m I_{z} & \m O_{z \times z} \\
        \m O_{z \times z} & \m O_{z \times z} \\
    \end{bmatrix}
     }^{\m E^r_s}
     \overbrace{
    \begin{bmatrix}
        \m y_s^{r}(k+1) \\
        \m \varphi_{out,s}^r(k+1)
    \end{bmatrix}
    }^{\m x^r_s(k+1)}
    = 
    \overbrace{
    \begin{bmatrix}
         \tilde{\m A}^r_1(k) & \m O & \dots & \m O & \m O \\
        \m O &  \tilde{\m A}^r_2(k)  & \dots & \m O & \m O \\
        \vdots & \vdots & \ddots & \vdots & \vdots \\
        \m O & \m O & \dots & \tilde{\m A}^r_s(k) & \m O \\ 
        \m O & \m O & \dots & \m O & \m -I_{z}\\
    \end{bmatrix}
    }^{\m A^r_s(k)}
    \overbrace{
    \begin{bmatrix}
        \m h^r_1(k) \\
        \m h^r_2(k) \\
        \vdots \\
        \m h^r_s(k) \\
        \m \varphi^r_{out,s}(k)
    \end{bmatrix}
    }^{\m x^r_s(k)}
    + \notag \\ &
    \overbrace{
    \begin{bmatrix}
        \tilde{\m B}^r_1(k) & \m O & \dots & \m O  \\
        \m O & \tilde{\m B}^r_2(k)  & \dots & \m O  \\
        \vdots & \vdots & \ddots & \vdots  \\
        \m O & \m O & \dots & \tilde{\m B}^r_s(k)  \\ 
        \m O & \m O & \m O & \m O
    \end{bmatrix}
    }^{\m B^r_s(k)}
    \overbrace{
    \begin{bmatrix}
        \m u^r_1(k) \\
        \m u^r_2(k) \\
        \vdots \\
        \m u^r_s(k) \\
    \end{bmatrix}
    }^{\m \sigma^r_s(k)}
    + 
    \overbrace{
        \begin{bmatrix}
        \tilde{\m \psi}^r_1(k) \\
        \tilde{\m \psi}^r_2(k) \\
        \vdots \\
        \tilde{\m \psi}^r_s(k) \\
        \m \sigma_s^r(k) \circ \m k_o^* \circ (\hat{\m h}_s^r(k))^{\circ 1/2}+ \m k_s^* \circ {\left(\max(\m y_s^r-\m p,0)\right)}^{\circ 3/2}
    \end{bmatrix}
    }^{\m \psi^r_s(k)}
\end{split}
\end{equation}}
{where $\m k_o^*$ and $\m k_s^*$ collects $ k_o$ and $ k_s$ for all reservoirs in all systems, respectively. Similarly, $\m p$ collects spillway elevations. Vectors $\m \varphi^r_{out}, \m y_s^r$, and $\m {\sigma}_s^r$ are defined as follows. Matrices $\m E_s^r $ and $ \m A_s^r \in \mathbb{R}^{2z \times 2z}$, matrix $\m B_s^r \in \mathbb{R}^{2z \times z}$, while $\m x_s^r \in \mathbb{R}^{2z}$, $\m \psi_s^r \in \mathbb{R}^{2z}$, and $\m \sigma_s^r \in \mathbb{R}^{z}$}. Vectors $\m \varphi_{out,s}^r, \m y_s^r, $ and $\m \sigma_s^r$ are defined as follows:
\begin{subequations}
\begin{align}
    \m \varphi^r_{out,s}(k) &= [\overbrace{\m q^{r,1}_{out,1}(k),\dots, \m q^{r,n_r}_{out,1}(k)}^{\m q^r_{out,1}(k)}, \dots , \overbrace{\m q_{out,s}^{r,1}(k) \dots \m q_{out,s}^{r,n_r}(k)}^{\m q^r_{out,s}(k)}]^T  \\
    \m y^r_s(k) &= [\overbrace{\m h^{r,1}_{1}(k),\dots, \m h^{r,n_r}_{1}(k)}^{\m h^r_1(k)}, \dots ,\overbrace{ \m h_{s}^{r,1}(k) \dots \m h_{s}^{r,n_r}(k)}^{\m h_s^r(k)}]^T  \\
    \m \sigma_s^r(k) &= [\overbrace{\m u^{r,1}_1(k) \dots \m u^{r,n_r}_1(k)}^{\m u_1^r(k)} \dots \overbrace{\m u^{r,1}_s(k) \dots \m u^{r,n_r}_s(k)}^{\m u^r_{s}(k)}]^T.
\end{align}
\end{subequations}
\hrule
\end{strip}
\subsection{1-D Channel Dynamics} ~\label{sec:channel}
The flow dynamics in rivers or channels can be modeled using hydrologic or hydraulic routing modeling approaches \cite{USArmyCorpsofEngineersInstituteforWaterResourcesHydrologicEngineeringCenter2000}. The first is governed by the water balances in the inlet and outlet sections of a sub-reach. The continuity equation and storage-outflow relationships are used to estimate the outflow in the last sub-reach \cite{mccarthy1938unit,cunge1969subject}. However, this approach is suitable only for estimating flows. Moreover, hydrologic routing is not flexible enough to explain more detailed phenomena such as backwater effects from downstream reservoirs or flood waves in channels with very flat slopes \cite{cunge1969subject}. Another issue is the time-related coefficients associated with the hydrologic routing equations. Typically, to ensure model's stability, coefficients are in the order of several hours and days. This poses as a drawback for real-time monitoring of urban channels \cite{kumar2011extended}. We propose a diffusive wave simplification in the Saint-Venant Equations (SVE) to represent the flow dynamics in the 1-D channels.

Recently, an application of the full one-dimensional Saint-Venant equations allowed the state-space representation  in channels \cite{bartos2021pipedream}. The authors developed a backward Euler implicit scheme solving continuity and momentum equations via a sparce matrix system of equations. While model stability is theoretically increased with this implicit numerical scheme, in order to develop a full state-space representation of watersheds, reservoirs, and channels, a implicit derivation for the other systems would be required. Therefore, we solve the 1-D flow dynamics using an Euler explicit numerical scheme as used in Eq.~\eqref{equ:state_watershed} and Eq.~\eqref{equ:state_space_reservoir}.
\subsubsection{Channel Water Depth Dynamics}
To represent the channel 1-D dynamics in space and time, we develop an explicit diffusive-wave simplification in SVE, assuming the friction slope from Manning's steady-flow equation. Fig.~\ref{fig:study_case} part c) shows a scheme of a 1-D channel. The outflow in each segment is calculated through Manning’s equation as follows \cite{Panday2004}:
\vspace{-0.1 cm}
\begin{equation}
    q_{out}^c(h^c_i(k)) = \frac{a_i^c(k) (r_{h,i}(k))^{2/3}}{n_i} \Bigl(\frac{\partial h^c_i(k)}{\partial y}\Bigr)^{1/2}
\end{equation}
where $a_i^c$ is the wetted area, $r_{h,i}$ is the hydraulic radius, $h^c_i$ is the channel water surface depth at cross section $i$, and $y$ is the channel's longitudinal direction. The previous expression is expanded in a vector format given by:
\begin{strip}
\vspace{-0.7cm}
\hrule
\begin{equation}\label{equ:manning_channel}
    \m q_{\m out}^c\left(\m h^{\m c}(k)\right) = {\ \m 1}_{n_c \times 1} \oslash \m n^c \circ \m a^c\left(\m h^c(k)\right)\circ \m (\m r_h^c\left(\m h^c(k)\right)^{\circ 2/3} {\ \circ 
    \left(\frac{\partial \m h^c(k)}{\partial \m y}\right)}^{\circ 1/2},
\end{equation}
\hrule
\vspace{-0.5cm}
\end{strip}
where $\m n^c$, $\m a^c$, $\m r_h^c$ , is the Manning's coefficient, cross section area function, and  hydraulic radius for each sub-reach.

A channel is discretized into sub-reaches with appropriate spatial resolution to be suitable for the time-step of the model \cite{chang2002courant}. Therefore, the model simulates steady non uniform flow assuming initial boundary conditions from $q_{out}^r(k)$ and from the outlet normal friction slope \cite{Panday2004,Maxwell2014}. The conservation of mass and momentum equations are given by: 
\begin{subequations} \label{equ:channel_diffusive}
\begin{align}
    \frac{\partial q_i^c(k)}{\partial y} &= - \frac{\partial a_i^c(k)}{\partial t} \label{equ:mass_conservation} \\
    \frac{\partial h^c_i(k)}{\partial y} &= s_0^c - \overbrace{\Bigl(\frac{q^c_i(k) n_i^c}{a_i(k) r_{h,i}(k)^{2/3}}\Bigr)^2}^{s_{f,i}(k)} \label{equ:momentum_conservation}
\end{align}
\end{subequations}
where $y$ represents the longitudinal channel dimension, $q_i^c$ is the net flow within cross sections, $s_0^c$ is the bottom slope, $n_i^c$ is Manning's roughness coefficient, and $s_{f,i}$ is the friction slope. 

\subsubsection{Channel Water Surface Depth Dynamics}

We can expand Eq.~\eqref{equ:mass_conservation}, resulting in a vectorized channel water surface depth mass balance as:
\begin{equation} \label{equ:channel_euler}
     \m h^c(k+1) = \m h^c(k) +  
     \Delta t\m 1_{n_c \times 1} \oslash (\Delta \m x \circ \Delta \m y ) \overbrace{\left(\m q_{in}^c(k) - \m q_{out}^c(k)\right)}^{\m \Delta \m q(\m h^c(k))},
\end{equation}
where $h_c$ is the water surface depth in each sub-reach of the channel,  the outflow $Q^c$ is given by the Manning’s equation that depends on known functions of cross section area and hydraulic radius for each reach of the channel and length ($\Delta \m y$) and width ($\Delta \m x$). For the first sub-reach, $q_{in}^c$ is equal the reservoir outflow $(q_{out}^r)$. Flows $q_{in,i}^c(k)$ and $q_{out,i}^c(k)$ are equal to $q^c_{i-1}$ and $q^c_{i}$, respectively. 

To explicit solve Eq.~\eqref{equ:channel_euler}, we need to derive an expression for $\Delta \m q^c(\m h^c(k))$. To this end, we apply the energy conservation within two consecutive cross sections neglecting the velocity head to develop a vector representation for the friction slope assuming normal depths, such that for a particular cell ($i$) in a 1-D channel domain, we can write:
\begin{equation} \label{equ:slopes}
    s_{f,i} = \frac{1}{\Delta y_i}\Bigl(e_{l,i} + h^c_i(k) 
    -e_{l,i+1} - 
    h^c_{i+1}(k))\Bigr)
\end{equation}
where $e_l$ is the bottom elevation of the sub-reach segment and $\frac{\partial h^c_i(k)}{\partial y} $ is the water surface slope in the cell $i$ at step $k$. 

Expanding the previous equation in a vector format, we can derive the momentum equation from Eq.~\eqref{equ:momentum_conservation} for all sub-reaches as a linear combination of $\m h^c(k)$ (see Eq.~\eqref{equ:manning_channel}, such that:
\vspace{-0.1 cm}
\begin{equation} \label{equ:slopes_matrices}
    \frac{\partial \m h^c(k)}{\partial \m y} = \m A_{ slope} \m h^c(k) + \m b_{ slope},
\end{equation}
\vspace{-0.1 cm}
where $\partial \m h^c/\partial \m y \in \mbb{R}^{n_c}$ is a vector representing the water slopes in each reach of the channel, $\m A_{slope} \in \mathbb{R}^{n_c \times n_c}$ and $\m b_{slope} \in \mathbb{R}^{n_c}$ are derived from Eq.~\eqref{equ:slopes} assuming 1-D connections between cells, computing the water surface slopes, and including the outlet slope boundary condition.

\subsubsection{1-D Flow in Open Channels}

Similarly, to the overland flow model for the watershed model, we can compute
$\left(\m q_{in}^c\left(k\right)- \m q_{out}^c\left(k\right)\right)$ as a function of a
flow direction matrix and the flow through Manning's equation. Therefore, the net flow $\Delta \m q^c(\m h^c(k)) = \m q_{in}^c(k) - \m q_{out}^c(k)$ for all sub-reaches can be given as
\begin{equation} \label{equ:direction_channel}
    \Delta \m q^c(\m h^c(k)) = \m B_d^{c} \m q_{out}^c(\m h^c(k)) + \m w(k),
\end{equation}
where $\m B_d^c$ is a topology matrix linking each sub-reach segment with the previous one and $\m w(k)$ is a zero vector where only the first entry is equal the reservoir outflow ($q_{out}^r(k)$).

However, to estimate the 1-D flow propagation in channels, we do not assume a simplified hydraulic radius and define a constant $\m \lambda$ due to the relatively small width of the sub-reaches when compared to the
gridded cells from the watershed system \cite{liu2004effect}. To estimate $\m \lambda$, functions that describe the cross section area ($\m a_{h}^c$(k)) and
hydraulic radius $(\m r_h^{c}(k))$ are required and can be derived in terms of topographic properties of the
channel. For a rectangular channel (i.e., condition of the case study), $\m a_h^c\left(\m h^c\right(k))=\Delta \m x \circ \m h^c(k)$, and
$\m r_h^c\left(\m h^c\right(k))=\m a_h^c(\m h^c(k)) \oslash  (\Delta \m x+\ 2\m h^c(k))$.
\subsubsection{Linearized Channel Dynamics}
We can substitute Eq.~\eqref{equ:direction_channel} into Eq.~\eqref{equ:channel_euler}, resulting in a matrix expression for the channel non-linear dynamics given by
\begin{strip}
\small
\hrule
\begin{equation} \label{equ:linearized_channel}
\begin{split}
    \m h^c(k+1) = \m h^c(k) + \Delta t\m 1_{n_c\times1} \oslash(\Delta \m x \circ \Delta \m y)\circ  \Bigl[ \m B_d^{c} \m 1_{n_c \times 1} \oslash \m n^c{\circ \m a}_h^c\left(\m h^c(k)\right)\circ \m r_h^c\left(\m h^c(k)\right)^{\circ 2/3}
    {\circ  \left(\m A_{slope} \m h^c\left(k\right)+\m b_{slope}\right)}^{\circ 1/2}) + \m w(k)\Bigr].
\end{split}
\end{equation}
\hrule
\end{strip}
\vspace{-0.8 cm}

Finally, we can obtain a linearized expression around an operation point $\m h_{0}^c$, applying the jacobian in the previous equation, such that
\begin{equation} \label{equ:gradient}
    \overbrace{\m I_{n_c}}^{\m E^c}\m h^c(k+1) = \overbrace{\left(\m I_{n_c} + \nabla \m h^c(\m h_{0}^c)\right)}^{\m A^c(k)} \m h^c(k),
\end{equation}
where $\nabla$ is the gradient of Eq.~\eqref{equ:linearized_channel} and can be obtained knowing the cross section area and hydraulic radius functions of each sub-reach and can be computationally derived with symbolic functions on MATLAB, for instance. The gradient and the operational point from Eq.~\eqref{equ:gradient} are refreshed each model time-step.

\subsection{State-Space Representation}
The derivation of the state-space model for a single watershed, reservoir, and channel is performed merging the dynamical state-space representation for each sub-system from Eqs.~\eqref{equ:state_watershed}, \eqref{equ:state_space_reservoir}, \eqref{equ:gradient}, and applying in Eq.~\eqref{equ:statespace}, the watershed-reservoir-channel dynamical system can be defined as
\begin{strip}
\vspace{-0.7cm}
\hrule
\begin{equation} \label{equ:state_space_final}
\hspace{-0.53cm}\overbrace{
\begin{bmatrix}
    \m E^w & \m O & \m O \\
    \m O & \m E^r & \m O \\
    \m O &  \m O & \m E^c
\end{bmatrix}
}^{\m E}
\overbrace{
\begin{bmatrix}
    \m x^w(k+1) \\
    \m x^r(k+1) \\
    \m x^c(k+1)
\end{bmatrix}
}^{\m x(k+1)}
= 
\overbrace{
\begin{bmatrix}
      \m A^w(k) & \m O_{(2q + 1) \times 1} & \m O_{(2q + 1) \times n_c} \\
      \m O_{1 \times (2q + 1)} & \m A^r(k) & \m O_{1 \times n_c} \\
      \m O_{n_c \times (2q + 1)} & \m O_{n_c \times 1} & \m A^c(k) \\
    \end{bmatrix}
}^{\m A(k)}    
\overbrace{
\begin{bmatrix}
    \m x^w(k) \\
    \m x^r(k) \\
    \m x^c(k) \\
\end{bmatrix}
}^{\m x(k)}
+  
      \overbrace{\m B^r(k)}^{\m B(k)} \m u(k) 
     + \overbrace{
\begin{bmatrix}
    \m \psi^w(k,\m x^w(k)) \\
    \m \psi^r(k, x^r_{*}, u^r_{*}) \\
    \m O_{n_c \times 1}
\end{bmatrix} 
}^{\m \psi(k)}
\end{equation}
\hrule
\vspace{-0.5cm}
\end{strip}

Similarly, the derivation of the state-space dynamics for the general case with multiple watersheds, more than one reservoir per watershed, and more than 1 channel per reservoir can be done using Eqs.~\eqref{equ:state_space_reservoir} and \eqref{equ:state_watershed}. It is noted that the watershed is an autonomous system, whereas the reservoir is governed by the control law, and the channel depends on the outflow from the reservoir, although is not directly controlled.
\section{Reactive and Predictive Control Strategies} ~\label{sec:static_controls}
The control strategies tested in this paper are focused on reducing flood effects. Here in this paper, we define it as a composite function accounting for local floods in reservoirs and channels, while minimizing rapid changes in valve operation. We can generally write the flood performance as a function defined for ($n_s$) systems accounting for flooding (e.g., reservoirs and channels) with different number of components per each system that impacts the flood performance ($c(i)$) (e.g., flood performance for reservoirs is associated with water levels and control deviations, whereas for channels is only associated with water levels). Therefore, we can generally write flood performance as linear combinations between weights and functions for each component of each system associated with floods, such that:
\begin{equation}
    \mathrm{Flood~Performance} = \sum_{i_s = 1}^{n_s} \sum_{j_c = 1}^{c(i)} \alpha_{i_s,j_c} f_{i_s,j}(\m X)
\end{equation}
where $\m X$ is the state matrix concatenating $\m x(k)$ for all simulation time.

The aforementioned equation is detailed defined in the objective function used to describe the flood mitigation performance in Sec.~\ref{sec:MPC}. This particular function operates in all modeled states returning a number a real number. Although the flood performance was defined here, only the Model Predictive Control strategy is able to control all parts of the system (e.g., channels and reservoirs) and act directly to maximize it, since the reactive controls are developed only using the reservoir dynamics. This section is organized as follows: first we describe the reactive control strategies tested in this paper and following that, we introduce a non-linear model predictive control algorithm to control valve operations in stormwater reservoirs. 

\subsection{Discrete Linear Quadratic Regulator}
Given a linearized time-invariant state space representation of the reservoir dynamics, the linear quadratic regulator strategy finds an optimal control signal for time step $k$ minimizing a quadratic cost function, solved as an unconstrained convex optimization problem given by
\begin{equation} \label{equ:LQR}
     \min_{\m u}{ J(\m u)}  = \sum_{n=1}^{\infty} \left[(\m h^r(k))^T\m Q \m h^r(k) + (\m u^r(k))^T \m R \m u^r(k)\right],
\end{equation}
where $\m Q$, and $\m R$ are tuned symmetric positive semi-definite matrices representing the setpoints, cost of the states, cost of controls and cross-term costs.

This particular control strategy only considers the reservoir dynamics. Therefore, the coupled flood performance of this case only focus on reservoir water depth and control energy as shown in aforementioned equation. During the period where the dynamics are unchanged (e.g., steady inflows from the watersheds), the performance of the LQR control is optimal \cite{Wong2018}. The solution of this equation given by the discrete algebraic Riccati equation (DARE) \cite{pappas1980numerical}, consists into finding a $\m P(k)$ matrix (cost-to-go) and a closed-loop feedback gain $\m K(k)$, such that 
\begin{equation} \label{equ:feedback_gain_riccati}
\small
\m K(k) = \Bigl(\m R + (\m B^r(k))^T \m P(k) \m B^r(k)\Bigr)^{-1} (\m B^r(k))^T \m P(k) \m A^r(k).
\end{equation}
where the aforementioned matrices $\m P(k)$ and $\m K(k)$ can be solved using the \textit{dlqr} function on Matlab.

\begin{figure*}
    \centering
    \includegraphics[scale = 0.2]{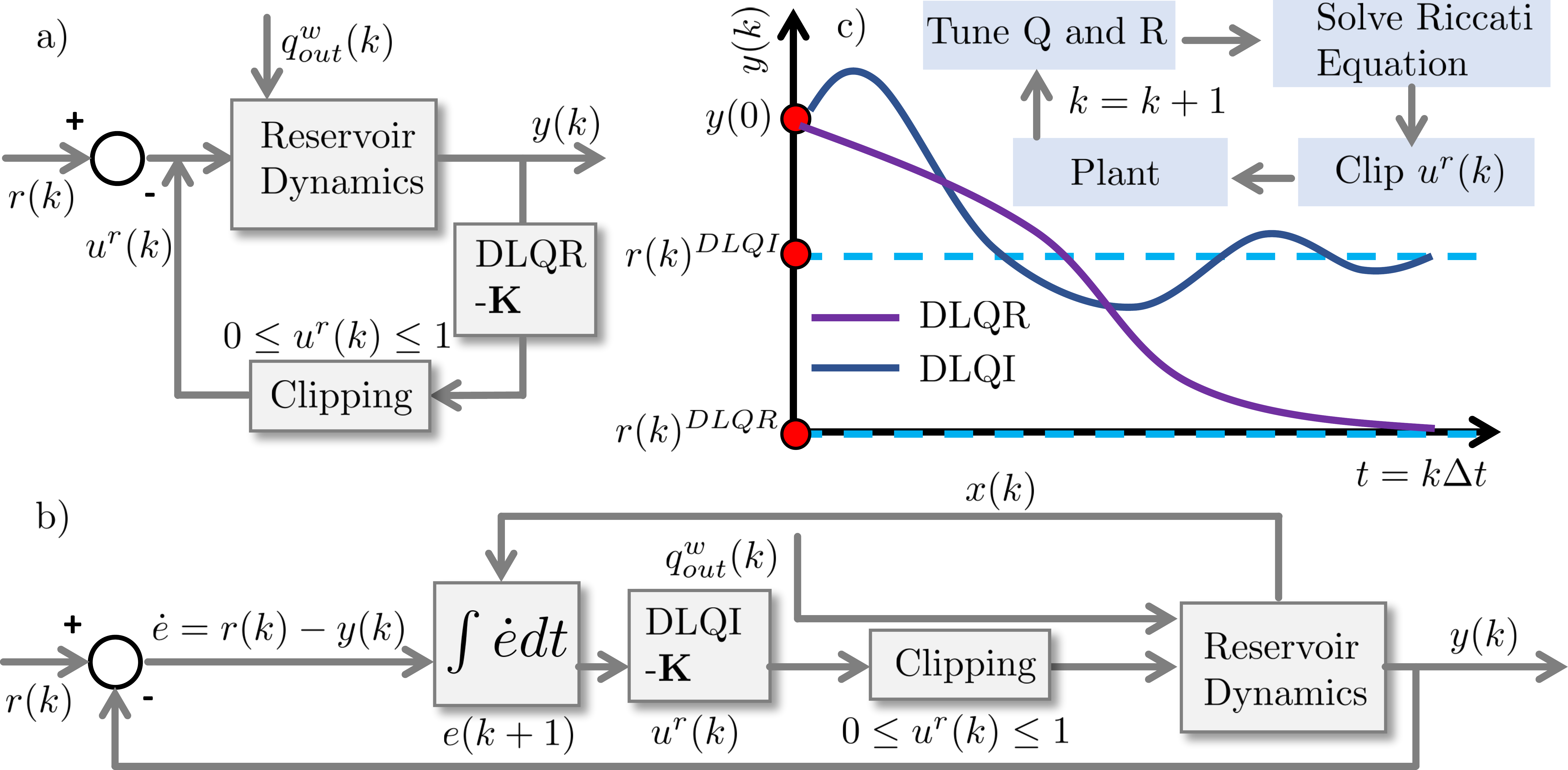}
    \caption{Controls based on the solution of Riccati Equation. Part a) represents the closed-loop system of the discretized linear quadratic regulator, Part b) is the block diagram of the discretized linear quadratic integrator with servo control, and part c) represents a schematic objective of the control approach.}
    \label{fig:LQR_chart}
\end{figure*}

A schematic detail of the closed-loop system for DLQR is presented in Fig.~\ref{fig:LQR_chart} part a). In this control algorithm, we are only using the reservoir plant in the dynamics. To design the state feed-back gain matrix $\m K$, we can first tune $\m Q=\m C^T \m C$ and $\m R=\m I_{n_r}$ as a first attempt. Basically, the DLQR control tends to find an optimal control for time $(k+1)$ that ultimately stabilize the reservoir dynamics or, in other words, tends to release the water considering the costs of rapid changes in the valves. To ensure physical constraints in the inputs \cite{Wong2018}, we limit the feed-back gain to 

A schematic detail of the closed-loop system for DLQR is presented in Fig.~\ref{fig:LQR_chart} part a).

\begin{subequations} \label{equ:u_bound}
\begin{align}
    \m u^r(k) &= -\m K \m h^r(k)  \\
     0 & \le   u^r_i(k) \le 1,~~i = \{1, 2, \dots, n_r\}.
\end{align}    
\end{subequations} 
\vspace{-0.5 cm}

\subsection{Discrete Linear Quadratic Integrator}
The linear quadratic integrator works similarly to the DLQR control; however, an augmented dynamical system to account for a reference tracking setpoint \cite{young1972approach} is developed. The system dynamics is augmented to include the temporal evolution of the integral of the error between the reference and the output. The setpoint reference can be a constant value or a time-varying function representing the goal of the control technique. Assuming the error between states and outputs as $\dot{\m e}(k) = \m r(k) - \m y(k)$ and discretizing it by an explicit forward Euler method in terms of the error integral ($\m e(k)$), the augmented dynamical system can be given as
\begin{strip}
\vspace{-0.5 cm}
\hrule
\begin{equation} \label{equ:LQI}
    \begin{bmatrix}
    \m h^r(k+1) \\
    \m e(k+1)
\end{bmatrix}
    = 
\begin{bmatrix}
    \m A^r(k) & \m O_{n_r\times n_r} \\
    - \m C^r & \m I_{n_r}
\end{bmatrix}
\begin{bmatrix}
    \m h^r(k) \\
    \m e(k)
\end{bmatrix}
+ 
\begin{bmatrix}
    \m B^r(k) \\
    \m O_{n_r\times n_r}
\end{bmatrix}
    \m u^r(k) +  \tilde{\m \psi}^r(k,\m x^r_{*}, \m u^r_{*}) + \Delta t
\begin{bmatrix}
    \m O_{n_r \times 1} \\
    \m 1_{n_r \times 1}
\end{bmatrix}
     \m r(k)
\end{equation}
\hrule
\vspace{-0.5 cm}
\end{strip}
where $\m r(k)$ is a setpoint or reference goal (e.g., minimum water surface depth to maintain natural ecosystems in the reservoir) and $\m C^r = I_{n_r}$.

The closed-loop system for the DLQI is presented in Fig.~\ref{fig:LQR_chart} part b). A known input given is by $q_{out}^w(k)$, which changes the dynamics over time. The solution of the DARE is performed with the coefficients of matrix  $\m Q$ (see Eq.~\eqref{equ:LQR} and Fig.~\ref{fig:LQR_chart}) representative of the deviations in states, outputs and from matrix $\m R$ control signal of $ 1 \times 10^0, 1.5\times 10^{3}$ and $1 \times 10^2$, respectively. The number of integrators is assumed to be equal the number of outputs, hence  all nodes are considered observable. After determining the optimal feedback gain, to ensure physical constraints, we also clip the control signal following Eq.~\eqref{equ:u_bound} as shown in Fig.~\ref{fig:Control_Analysis} part b). The control input can be separated into
\begin{equation} \label{equ:LQI_nodes}
     \m u^r(k) = -
    \begin{bmatrix}
        \m  k_s & \m  k_e
    \end{bmatrix}
    \begin{bmatrix}
        \m  h^r(k) \\
        \m  e(k)
    \end{bmatrix},
\end{equation}
where $\m k_s$ is the system feedback gain and $\m  k_e$ the servo feedback gain.

The ultimate objectives of DLQR and DLQI are presented in Fig.~\ref{fig:LQR_chart} part c). While DLQR tends to find a control schedule steering the output function to zero (e.g., rapidly release stored water and decrease $h^r(k)$ considering control energy), DLQI does a similar approach but for a reference water level $r(k)^{DLQI}$ (i.e., a given expected reservoir water level). 
\subsection{Ruled-Based Controls}
Typically, stormwater systems are passive or controlled by ruled-based reactive and local controls, sometimes even manually performed by operators \cite{Schmitt2020,Shishegar2019,Garcia2015}. Although the control rule of these methods seem logical and easily applicable for simple storm events, they can fail to control more complex cases as fast consecutive storms by not predicting the future behavior of the system \cite{Lund2018,Sharior2019}. Some of the most common rules for reservoirs for flood control are presented in Table~\ref{tab:ruled_based} and tested in this paper. Some control decisions are made for these type of heuristic controls, such as the critical water surface depth to open the valves for the on-off approach and the required retention time after a storm event to start releasing water for the detention control approach. These parameters were assumed as $h_{cr}$ = 3 m (i.e., representing a critical level in terms of flood management) and $t_d$ = 6 hrs (i.e., representing a minimum detention time for sedimentation).

\begin{table}
\footnotesize	
\caption{Types of static ruled-based controls} ~\label{tab:ruled_based}
\centering
\begin{tabular}{ll} 
\hline
Type                                   & Description                                  \\ 
\hline\hline
Passive Control (a)                    & Valve always 100\% open                      \\ 
\hline
\multirow{3}{*}{Detention Control (b)} & If an event occurs, valve opening = 0\%      \\
                                       & After the event, valve opening = 0\% for $t_d$  \\
                                       & Else, valve opening = 100\%                  \\ 
\hline
\multirow{3}{*}{On/Off (c)}            & If $h < h_{cr}$, valve opening = 0\%         \\
                                       & If $h \ge h_{cr}$, valve opening = 100\%     \\
                                       & Else, valve opening = 0\%                    \\
\hline\hline
\end{tabular}
\end{table}

\subsection{Model Predictive Control} ~\label{sec:MPC}
For a given time, the Model Predictive Control strategy estimates the future behavior of the system, finds the solution of an optimization problem and defines a control trajectory. The control strategy is estimated with actual and predicted states. As a result, a simplified plant of the complete dynamical system (e.g., a simplified version of the Shallow Water Equations for 1-D and 2-D overland and channel flow) is utilized. The use of a simplified plant is typically a requirement in order to have a relatively fast model that could be run in real-time. The MPC approach can be time-consuming due to consecutive optimizations, therefore defining the proper model simplifications in the plant while maintaining accurate solutions is a challenge \cite{Lund2018}. 

\begin{figure*}
    \centering
    \includegraphics[scale = 0.2]{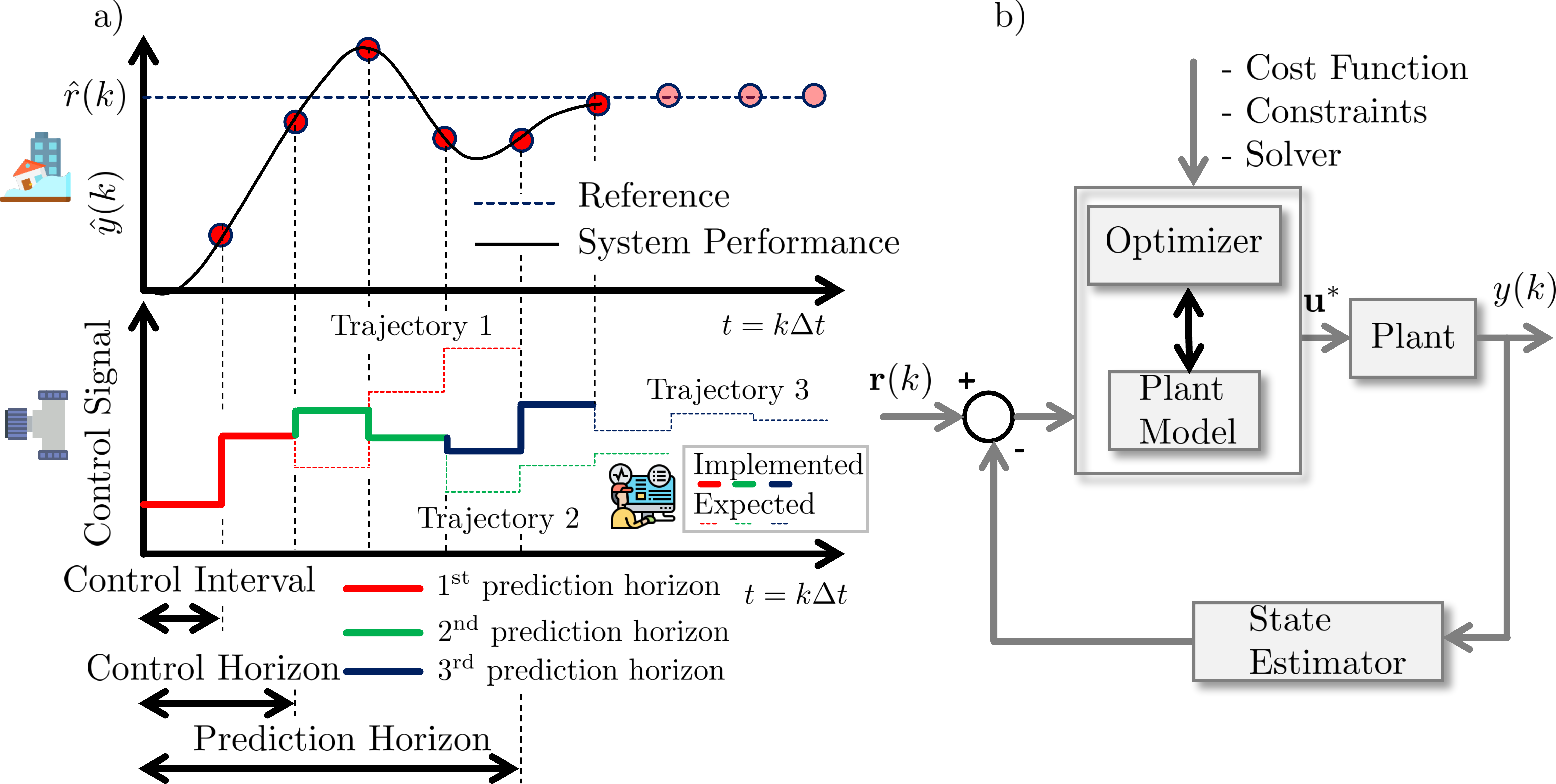}
    \caption{MPC schematic (a) and block diagram (b), where trajectories are generated for each moving prediction horizon, controls are assumed for every control horizon and are considered a constant piecewise within the control interval }
    \label{fig:MPC_chart}
\end{figure*}

Moreover, due to the nature of hydraulic and hydrological systems, the selection of the optimizer solver is also important. We test state-of-the-art optimizers as Global Search \cite{kearfott2013rigorous}, and Genetic Algorithms \cite{giacomoni2017multi,vose1999simple}, and the Interior-Point method was selected due to its robustness and faster computation. Therefore, in this paper, we develop a non-linear MPC solved with the Interior-Point method \cite{potra2000interior}. A flowchart of the MPC control algorithm is presented in Fig.~\ref{fig:MPC_chart}. Generally, an optimal control trajectory is solved for a prediction horizon, although only a few of the control signals are implemented in real-time (see red line in the control signal chart at Fig.~\ref{fig:MPC_chart}. Following that, a new optimization problem is solved resulting in the green line (see Fig.~\ref{fig:MPC_chart}). The solution of the MPC optimization problem is an optimal control trajectory vector $\m U_k=[u_1,  u_2\dots{} u_{N_p - 1}]^T$ such that the following optimization problem is satisfied:

\begin{strip}
\vspace{-0.7 cm}
\hrule
\begin{equation} \label{equ:optimization_problem}
    \min_{\m U_k}{\sum_{k=0}^{N_p-1} \m J(\m x(k+1),\m u(k+1))} = \  
     \rho_u \Delta \m  U_k^T \Delta \m U_k+ \rho_x \Delta \m H_k^T \Delta \m H + \rho_r\left(\max{\left(\m H_k- \m h_{ref}^{r},0\right)}\right) \notag  +  \rho_c\left(\max{\left(\m H^c_k- \m h_{ref}^c,0\right)}\right)
\end{equation}
\begin{subequations} ~ \label{equ:constraints}
	\begin{align}
\subjectto \;		\m E \m x(k+1)&=\m A(k)\m x(k) + \m B(k) \m u(k) + \m \psi(\m x(k), \m x^r_{*},\m u^r_{*}) \\
		\Delta{}u_{min}&\leq{}\Delta u\left(k\right)\leq \Delta{}u_{max} \\
		\m y\left(k\right)&=\m C \m x(k) \\
		\m u(k)&\in{} \mathcal{\m U} \\
		\m x(k)&\in{} \mathcal{\m X}
	\end{align}
\end{subequations}
\hrule
\vspace{-0.5 cm}
\end{strip}
where $\m J$ is the cost function (e.g., a function where weights are given for the control input $(\rho_u)$, for the water surface depths in the reservoirs $(\rho_r)$ and water surface depth in the channels $\rho_{c}$. $N_p$ is the prediction horizon, $\Delta u(k) = u(k) - u(k-1)$, $\Delta \m U_k = [\Delta u_1 \dots \Delta u_{N_p-1}]^T$ $\m H_k = [h^r_1, \dots h^r_{N_p - 1}]^T$, $\m H^c = [\m h^c_1, \dots \m h^c_{N_p - 1}]^T$ and $\mc U$ and $\mathcal{\m x}$ are the feasible sets of the control signal (i.e., $u(k) \in [0;1]$ and $\m x(k) \ge \m 0$). This objective function penalizes control schedules that generate states violating the threshold values for the maximum water surface depth in the channel and in the reservoir, while minimizing control energy. 

\section{Mathematical Model Application} ~\label{sec:Model_Application}
This section describes the numerical case studies that we apply the modeling approach. All codes are available in a open repository \cite{RTC_Code}. We attempt to answer the following questions:

\begin{itemize}
    \item Q1: \textit{How do control strategies primarily focused on increasing detention times affect flood control?}
    \item Q2: \textit{How do state-of-the-art reactive controls based on real-time measurements perform against a predictive control based on rainfall forecasting?} 
    \item Q3: \textit{Do reactive controls perform better than passive control?}
    \item Q4: \textit{Does the decrease in the control interval compensate the lack of predictability of reactive controls compared to a predictive control?}
\end{itemize}

To enable feasibility to use the controls in real world, we assume a control interval of 15-min for the reactive controls. Therefore, the control signal is assumed as a continuous piecewise within two intervals of 15-min. For the predictive control strategies, we assume a control interval of 60-min.

\begin{figure*}
    \centering
    \includegraphics[scale = 0.2]{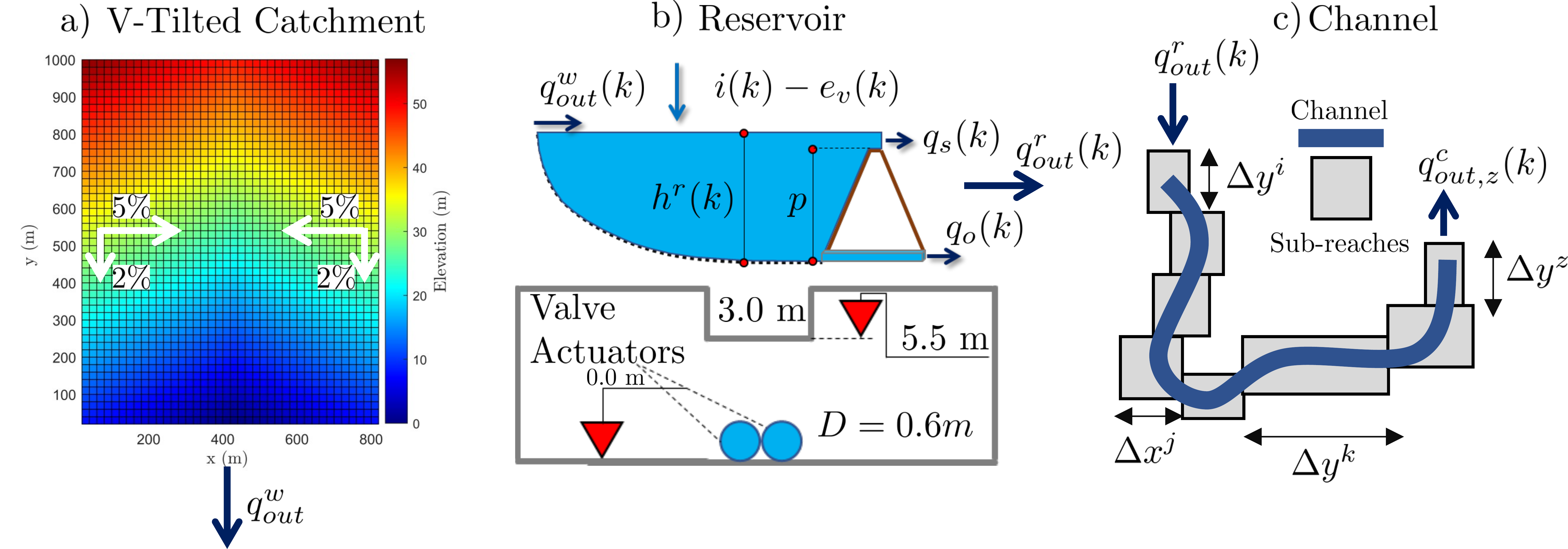}
    \caption{System composed by an autonomous watershed, a controlled reservoir and an autonomous channel receiving flow from the reservoir, where a) represents the watershed, (b) the reservoir, including spillway and orifice information, and c) a schematic representation of the 1-D channel.}
    \label{fig:study_case}
\end{figure*}

We test the effects of RTC in a system defined by a watershed (a), reservoir (b) and channel (c), as shown in Fig.~\ref{fig:study_case}. The V-Tilted catchment has only one cell as the outlet ($i_0$ and $j_0 \in \mathbb{R}$) and dimensions of $1.62$ km $\times 1$ km and is composed by two rectangular planes of $0.8$ km $\times 1$ km m and a channel of $ 20$ m  in the connection within the planes \cite{Fry2018,Maxwell2014,Panday2004}. The $x-x$ slopes in the hillslopes is $5\%$ and the $y-y$ slope in the channel longitudinal direction is $2\%$. Two types of land cover are assumed, one for the hillslopes representing the watershed and another for the channel representing the main open channel drainage collector. The reservoir area is defined as $0.65\%$ of the catchment area. The time step is assumed as $1 $ sec to ensure model numerical stability. All parameters of the model for each system are presented in the Table~\ref{tab:initial_parameters}. 

\begin{table*}
\footnotesize
\caption{Parameters required in the numerical case studies, where * represents changed values for the continuous simulation analysis} ~\label{tab:initial_parameters}
\centering
\begin{tabular}{lllll} 
\hline
System                      & Parameter     & Value  & Unit                                         & Description                       \\ 
\hline\hline
\multirow{13}{*}{Watershed} & $R_w$                      & 20     & m                                            & Cell Resolution                   \\
                            & $n_{per}$                  & 0.3, $0.18^*$    & s.m\textsuperscript{1/3}                     & Manning's coefficient             \\
                            & $ \zeta_{per}$              & 110, $0.6^*$    & mm                                           & Suction Head                      \\
                            & $\Delta \theta_{per}$      & 0.453, $0.386^*$  & cm\textsuperscript{3}.cm\textsuperscript{-3} & Moisture Deficit                  \\
                            & $k_{sat,per}$              & 10.92, $1.2^*$ & mm.h\textsuperscript{-1}                     & Saturated Hydraulic Conductivity  \\
                            & $h_{0,per}$                      & 10, $10^*$ & mm                                                     & Initial Abstraction  \\
                            & $h_{0,imp}$                      & 0, $0^*$ & mm                                                     & Initial Abstraction  \\   
                            & $F_{d_0,imp}$                      & 10, $10^*$ & mm                                                     & Initial Infiltrated Volume  \\                          
                            & $n_{imp}$                  & 0.018  & s.m\textsuperscript{1/3}                     & Manning's coefficient             \\
                            & $e_t$                      & 2  & mm.day\textsuperscript{-1}                      & Daily Evapotranspiration Rate             \\                          
                            & Hillslopes                 & 5      & \%                                           & Slopes in x direction             \\
                            & Channel Slope              & 2      & \%                                           & Slopes in y direction             \\
                            & $n_{cells}$                & 4050   & -                                            & Number of cells                   \\
                            & $A_d$                      & 1.62   & km\textsuperscript{2}                        & Drainage Area                     \\ 
\hline
\multirow{5}{*}{Reservoir}  & $\omega^r(h^r)$         & 10,530  & m\textsuperscript{2}                         & Reservoir Area                    \\
                            & $k_o$         & 1.538  & m\textsuperscript{1/2}s\textsuperscript{-1}  & Orifice coefficient               \\
                            & $k_s$         & 6.3    & m                                            & Spillway coefficient                  \\
                            & $h_s$         & 5.5    & m                                            & Spillway elevation                    \\
                            & $\eta$             & 1      & -                                            & Reservoir average porosity        \\ 
\hline
\multirow{7}{*}{Channel}    & $\Delta x$            & 3      & m                                            & Channel width                     \\
                            & $\Delta y$            & 30     & m                                            & Sub-reach length                      \\
                            & $n$             & 0.3    & s.m\textsuperscript{1/3}                     & Manning's coefficient             \\
                            & Sub-reaches   & 100    & -                                            & Number of reaches                 \\
                            & $s_0$         & 2.5    & \%                                           & Channel bottom slopes             \\
                            & $\frac{\partial h^c_o}{\partial y}$  & 2.5    & \%                                           & Outlet Boundary Condition                      \\
\hline\hline
\end{tabular}
\end{table*}

\subsection{Numerical Case Study 1: Consecutive Design Storms}
Typically, stormwater reservoirs are designed to hold rare storms such as 50 or 100-yr storms. However, two consecutive more recurrent storms can provide more flood risks in cases where the soil is likely more saturated and when control strategies focused on water quality by increasing retention times are implemented. Therefore, we test the efficiency of the control strategies for a 25-yr followed by a 10-yr design storm using a fitted Sherman intensity-duration-frequency curve for the available data for San Antonio - Texas \cite{atlas14precipitation}. The consecutive rainfalls have volumes of 7.5 and 6.3 inches, respectively. The landuse and landcover assumed for the watershed is equivalently to a recharge zone, with a relatively high infiltration capacity and can also be considered as a shrub forest for the hillslopes \cite{sharif2013physically,rawls1983green,Rossman2010}. For the main channel, however, a concrete Manning's coefficient was considered. Using a moving prediction horizon of 8 hours, control horizons of 2 hours and control intervals of 1 hour, the performance of the predictive optimization-based control algorithm is compared to reactive approaches.

\subsection{Numerical Case Study 2 - Continuous Simulation}
In order to assess the trade-offs between flood mitigation and control efforts, we test the mathematical model through a real observed rainfall time series that occurred between 04/22/2021 and 07/22/2021 in San Antonio Texas \cite{Survey2021}. The cumulative rainfall in this period exceeds the average 30-yr rainfall pattern and can be considered an intense rainy season \cite{GreggEckhardt2021}. Moreover, we also test a change in the land cover from a recharge zone to a clay soil \cite{furl2018assessment}. The hillslopes roughness were also changed to represent urban areas. Using a moving prediction horizon of 2 hours (e.g., most radars have a good precision within this interval) \cite{Lund2018}, control horizons of 2 hours and control intervals of 1 hour, the performance of the optimization-based control algorithm is compared to static-rules approaches.

\subsection{Performance Evaluation}

The metrics used in this paper to evaluate the control performance were the (a) Peak Flow Reduction $\eta_p$, (b) Relative Maximum Flood Depth $\eta$, (c) Relative Control Effort $\eta_c$, and (d) Flood Duration $\eta_d$. Each metric is applied for all control methods with $m$ representing the index of the control strategy. Given a duration from a initial time-step $k_b$ and a final time-step $k_f$, the duration $\Delta k = k_f - k_b$ are defined to capture a particular important time (e.g., maximum outflow peak). The concatenated states at this interval are vectors with $\Delta k$ rows for outflows and controls and matrices with $\Delta k$ rows for the channel water depth. We denote the symbol \dag~for concatenated states spanned over time. Therefore, the evaluation metrics are represented as follows:  
\begin{equation} \label{equ:peakflow}
\eta_{p,m} = \frac{\max (\m q_{out}^{w,\dag}(k_b:k_f)) - \max (\m q_{out,m}^{r,\dag}(k_b:k_f))}{\max (\m q_{out}^{w,\dag}(k_b:k_f))}, 
\end{equation}
\begin{equation} \label{equ:relativedepth}
\small
\eta_{r,m} = \frac{\max (\m H^{c,\dag}_m(k_b:k_f))}{h^c_{lim}},
\end{equation}
\begin{equation} \label{equ:relativecontrol}
\small
\eta_{c,m}(k) = \frac{\sum_{i=k_b}^{k_f}{\Delta u^{r,\dag}_m(i)}}{\max\limits_{m \in \mathfrak C}\sum_{i=k_b}^{k_f}{\Delta u^r(i)}},
\end{equation}
\begin{equation} \label{equ:floodduration}
\small
\eta_{d,m}(k) = \textbf n\bigl(\bigcup \limits_{t \in \mathfrak T}(\m h^\dag_{c,m} > h_c^{lim})\bigr) \Delta t,
\end{equation}
where $h^c_{lim}$ is the maximum allowable water surface depth in the channel, $\mathfrak C$ is the set of assessed controls, $m$ is the assessed control, $\mathfrak{T}$ is the duration set of the analysis. Other similar metrics applied to real-time controls can be found in \cite{Wong2018,Shishegar2019,Schmitt2020}.
\section{Results and Discussion} ~\label{sec:Results}
This section presents results and discussions on how MPC can increase flood performance compared to other reactive control strategies. First, we assess the control performance to critical consecutive design storms in San Antonio in Sec.~\ref{sec:design_storms}. Following that, we perform a continuous simulation with observed data and assess the performance of MPC in contrast to reactive controls in Sec.~\ref{sec:continuous_simulation}
\subsection{Design Storms} \label{sec:design_storms}
\begin{figure*}
    \centering
    \includegraphics[scale = 0.7]{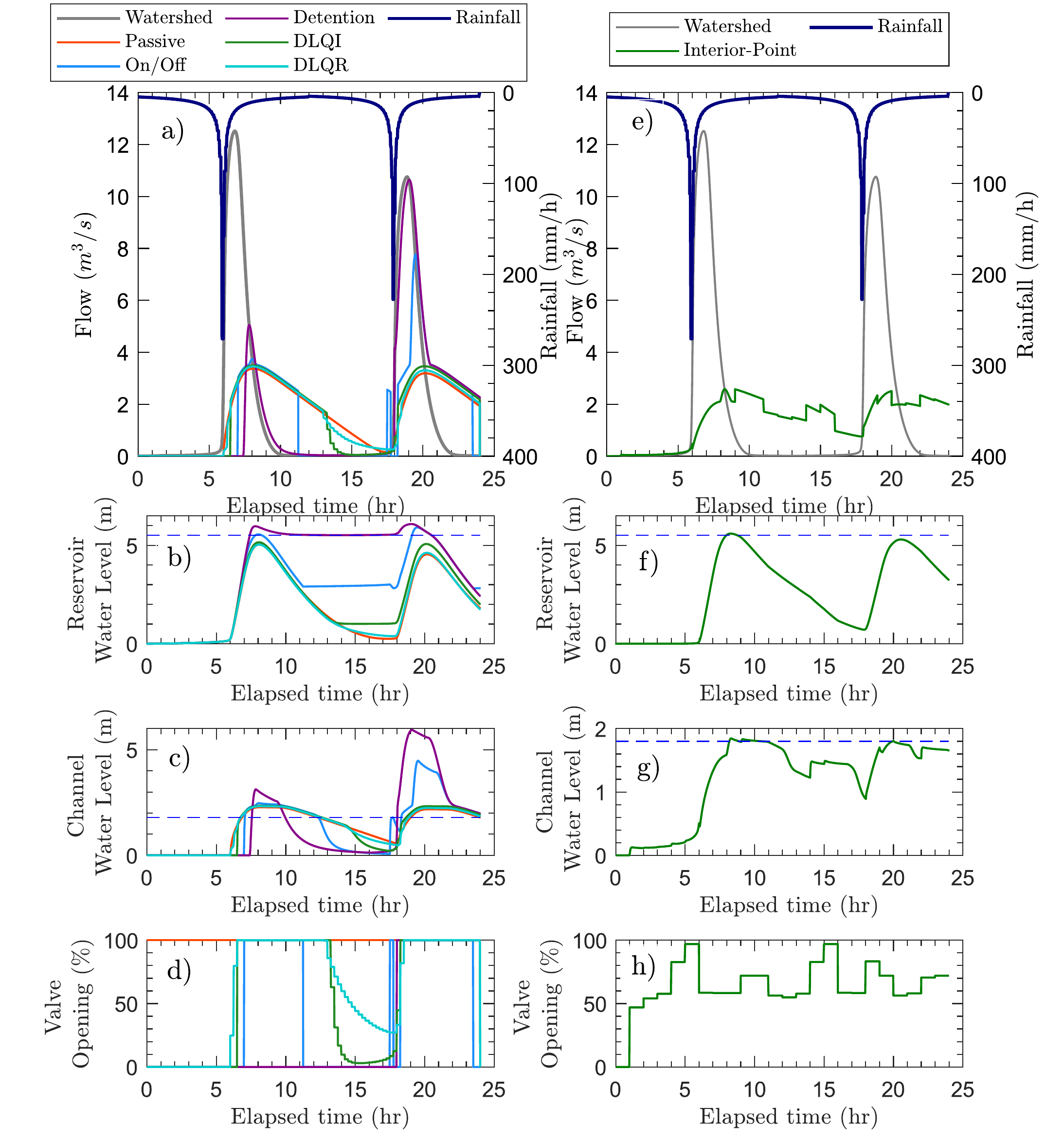}
    \caption{Control performance of ruled-based approaches in comparison with optimization-based control using the Interior-Point method, where the dashed blue lines in b) and f) represent the spillway elevation and in c) and g) they represent the allowed water surface depth in the channel for a 25-yr, 12-hr storm followed by a 10-yr, 12-hr storm.}
    \label{fig:Control_Analysis}
\end{figure*}

The comparison between reactive and predictive control approaches is shown in Fig.~\ref{fig:Control_Analysis}. The events produced similar peaks due to the initial saturation provided from the first storm. Outflow peaks in the reservoir increased from the first to the second storm, even though the runoff from the catchment decreased. The performance summary of ruled-based control against model predictive control is shown in Table~\ref{tab:comparison_controls}. The control approach that generated the highest outflow peak was the detention control. This approach basically started to release the stored water at 18 hours from the beginning of the first event (i.e., 6-hours from the end of the first storm event) and it matched with the start of the intense inflow runoff volume at the detention basin originated from the second storm. Only the detention control and the on/off strategy reached the spillway elevation and hence had the higher outflows. For the MPC control, smaller flows than all RBCs are observed. From these results it is noted that reactive controls, especially the ones primarily designed for water quality purposes, can increase flood and consequently erosion downstream for large storm events \cite{Schmitt2020}.

The detention pond dumps faster on the passive control strategy as expected. However the temporal evolution of the water surface depth in the reservoir had a similar behavior compared to DLQI and DLQR controls. It suggests that using reactive controls for relatively large floods might be as effective as the passive control. For the DLQI is noted that the state $h^r(k)$ is steered to the reference setpoint $ (r(k) = 1$ m$)$  where the system dynamics is not changing dramatically (i.e., cases where the inflow is steady). This type of control strategy is more suitable where constraints of minimum water depths are required in the reservoir (e.g., wetlands). Significant differences between the controls approaches; however, are depicted in the channel stage variation over time. For all reactive controls, periods of flooding occurred whereas the MPC control largely avoided it, although we assumed a perfect 8-hour rainfall forecasting.

For the MPC, a solution of a non-linear optimization problem is developed for each control horizon. We used the interior-point method, which is a gradient-based method dependent on the initial estimated control decision. Therefore, to initialize the optimization algorithm, we assume this initial point as a random combination from the previous control signal. Basically, a normal distributed random number with 0 average and 0.2 variance is generated $(\delta = \sim \mathcal{\m N}(0,0.2))$, and applied to the previous control, such that the initial point for the optimization is $u_0(k+1) = u(k)(1 + \delta)$. Defining this initial estimate; however, does not imply solutions only within this interval but actually can substantially decrease the computational time by starting in initial points nearby the previous control state.
\vspace{-0.5 cm}
\begin{table*} 
\footnotesize
\caption{Comparison within control strategies for a consecutive 25-yr, 12-hr and 10-yr, 12-hr design storm in San Antonio for a watershed in a recharge zone} ~\label{tab:comparison_controls}
\centering
\begin{tabular}{ccccccc} 
\hline
\begin{tabular}[c]{@{}c@{}}Type of \\Control\end{tabular}                    & \begin{tabular}[c]{@{}c@{}}Control \\Strategy\end{tabular}         & \begin{tabular}[c]{@{}c@{}}Peak Flow \\Reduction \\1\textsuperscript{st} peak (\%)\end{tabular} & \begin{tabular}[c]{@{}c@{}}Peak Flow \\Reduction \\2\textsuperscript{nd} peak (\%)\end{tabular} & \begin{tabular}[c]{@{}c@{}}Relative {}Maximum\\Flood \\Depth (\%)\end{tabular} & \begin{tabular}[c]{@{}c@{}}Relative \\Control \\Effort (\%)\end{tabular} & \begin{tabular}[c]{@{}c@{}}Flood {}Duration \\(hr)\end{tabular}  \\ 
\hline\hline
\multirow{5}{*}{\begin{tabular}[c]{@{}c@{}}Static / \\Reactive\end{tabular}} & Passive                                                            & 73.08                                                                                           & 68.67                                                                                           & 26.57                                                                                          & 0.0                                                                      & 10.40                                                                            \\
                                                                             & On/Off                                                             & 37.97                                                                                           & 27.81                                                                                           & 148.89                                                                                         & 93.46                                                                    & 10.98                                                                            \\
                                                                             & \begin{tabular}[c]{@{}c@{}}Detention \\Control\end{tabular}        & 15.01                                                                                           & 1.09                                                                                            & 229.87                                                                                         & 15.58                                                                    & 8.38                                                                             \\
                                                                             & DLQI                                                               & 72.11                                                                                           & 67.54                                                                                           & 30.18                                                                                          & 61.37                                                                    & 11.15                                                                            \\
                                                                             & DLQR                                                               & 72.42                                                                                           & 67.91                                                                                           & 29.01                                                                                          & 53.74                                                                    & 10.83                                                                            \\ 
\hline
\begin{tabular}[c]{@{}c@{}}Optimization-\\based / \\Predictive\end{tabular}  & \begin{tabular}[c]{@{}c@{}}Interior-Point \\Optimizer\end{tabular} & 79.39                                                                                           & 76.02                                                                                           & 2.55                                                                                           & 48.91                                                                    & 2.2                                                                              \\
\hline\hline
\end{tabular}
\end{table*}
\vspace{0.4 cm}
\subsection{Continuous Simulation} \label{sec:continuous_simulation}
\begin{figure*}
    \centering
    \includegraphics[scale = 0.6]{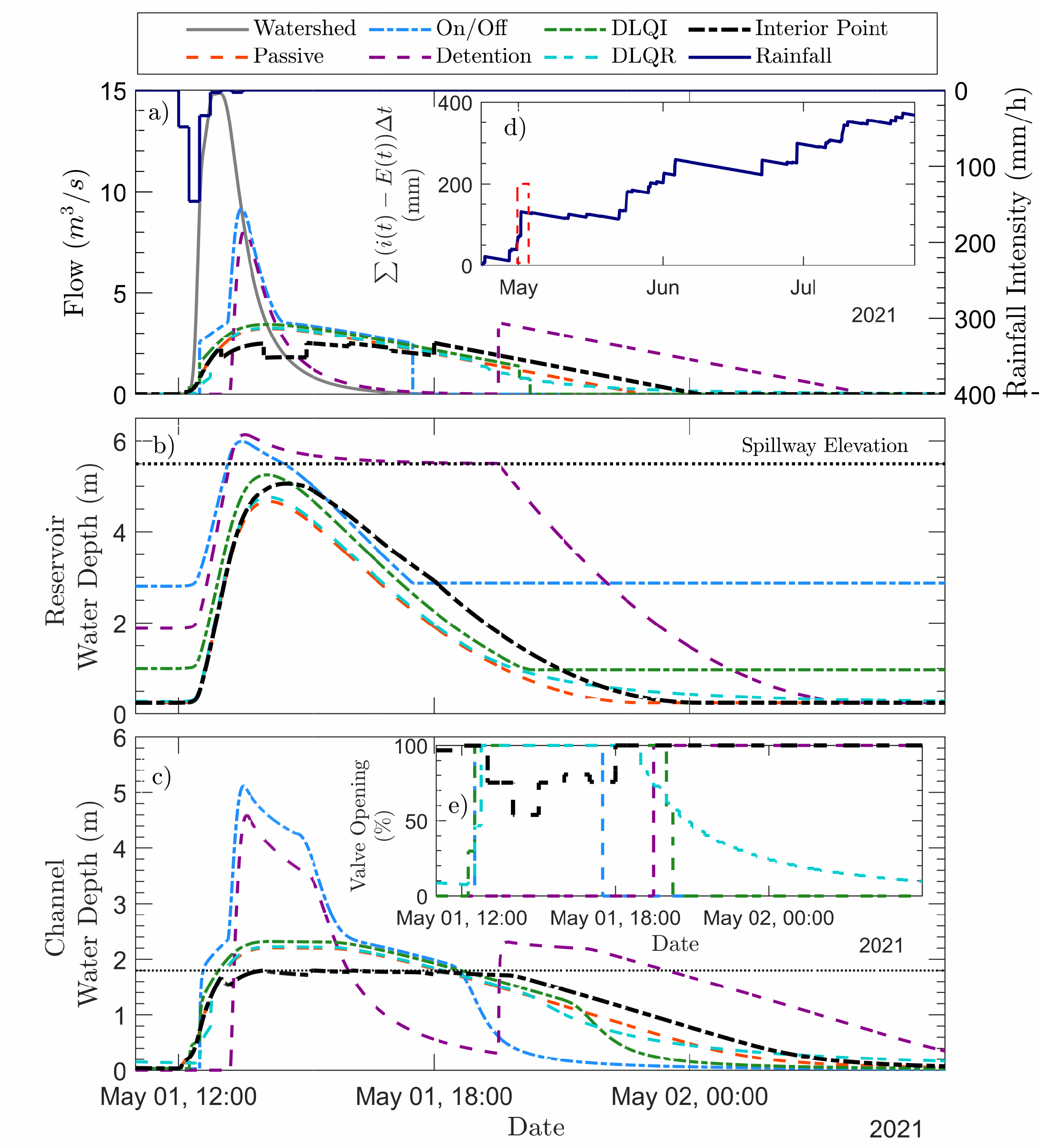}
    \caption{Control Performance of ruled-based approaches in comparison with optimization-based control using the Interior-Point method for the rainy season of 2021 in San Antonio, where the dotted black lines in b) and c) represent the spillway elevation and maximum allowable water surface depth in the channel, respectively. Parts d) and e) represent the net rainfall and control schedule of the assessed control algorithms, respectively.}
    \label{fig:Analysis_2months}
\end{figure*}

\begin{figure}
    \centering
    \includegraphics[scale = 0.54]{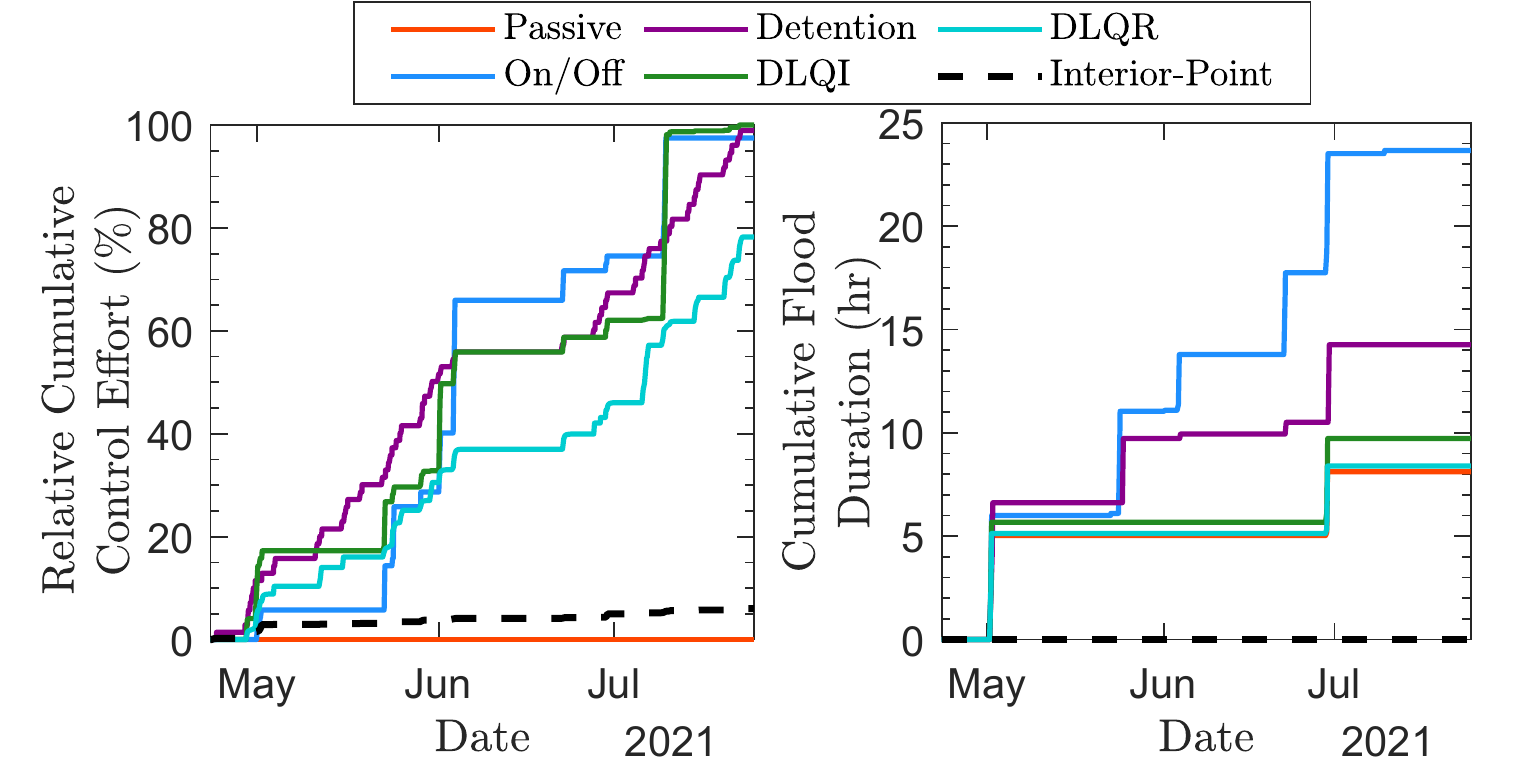}
    \caption{Trade-offs between control effort and flood duration for a rainy season in an urbanized watershed with poor infiltration capacity}
    \label{fig:Control_Evolution}
\end{figure}

The comparison between reactive approaches with an optimized-based/predictive control approach in a continuous simulation is shown in Fig.~\ref{fig:Analysis_2months} and the trade-offs between control effort and flood duration for the continuous modeling are shown in Fig.~\ref{fig:Control_Evolution}. In this analysis, we assume a 2-hour rainfall forecasting, showing the performance of RTC for low degrees of uncertainty in rainfall. In Fig.~\ref{fig:Analysis_2months}, a detail of the storm event that occurred in May 1\textsuperscript{st} of 2021 shows the hydrographs and the water surface depths in the reservoir and channels, as well as the control schedules over time. 

This rainfall event was rapid and intense, producing a high inflow peak due to the initial saturation of the soil from previous rainfall events. The hydrograph shows that the on/off and detention control had the highest outflows peak. The valves were mostly closed for all reactive control strategies in the arriving of the inflow peak since no relatively high water stages were yet observed. However, the on/off started this event with approximately 3 m of water stage (i.e., its control strategy is regulated by this water depth). 

While almost all other control strategies (i.e., except detention control) were able to release the accumulated volume with reasonable maximum channel water depths, the On/Off strategy, however, produced the highest outflow peak. It occurred due to opening both orifices lately without avoiding spillway outflow. The risk of flooding is generally increased with this strategy. Although the detention control had some of the inflow volume spilled, all the stored volume correspondent to approximately 2.5 m (i.e., 5.5 m - 3.0 m) of the pond stored volume (see Fig.~\ref{fig:Analysis_2months} part b) was successfully released 6 hours following the end of the rainfall event. As shown in Sec.~\ref{sec:design_storms}, this available time might not always be feasible if a new storm event occur in this interval. 

The DLQI, DLQR and Passive control strategies had  had similar outflows. The common reaction of the regulators is to stabilize the changing dynamics described by $\m \psi (k,\m x(k),\m u(k))$ due to $q_{out}^w(k)$ (see Fig.~\ref{fig:LQR_chart}), therefore, rapid valve openings is a common solution chosen by these controls depending on the tunned matrices $Q$ and $R$. One also can tune these matrices differently to favorably change the relative importance of variations in control schedule and water stage over time.

It is noted from Fig.~\ref{fig:Analysis_2months} that the the detention control and the on/off control had the highest stored water surface depths, which is in accordance with \cite{Sharior2019}. The latter control; however, returned for the regulated water surface depth relatively fast due to decide to open the valves after reaching 3 m whereas the detention control only released the flow hours afterwards. Moreover, even after the event, some volume is still stored for all controls. Outflow discharge only occurs when $h^r(k) > \max({h_o,0.2 \times D_h})$, where $D_h$ is the hydraulic diameter. Thus, the minimum water surface depth is approximately 0.24 m. 

An interesting result is the fact that the predictive control had a relatively larger stored water surface depth in the reservoir compared to the other controls. Despite this fact, it still  regulated the maximum water surface depth in the channel below the reference $h_{ref}$ of 1.8 m, as shown in Fig.~\ref{fig:Analysis_2months} c). During the peak of the storms, the predictive control decided to partially close the valves, which was a feasible decision because the water surface depth in the pond was below the spillway elevation and no intense inflow was expected within the next few hours. This type of decision is counter intuitive and pinpoints the benefit of using predictive optimization-based control algorithms rather than solely expertise-based manually operations in real-time. 

\begin{figure*}
    \centering
    \includegraphics[scale = 0.65]{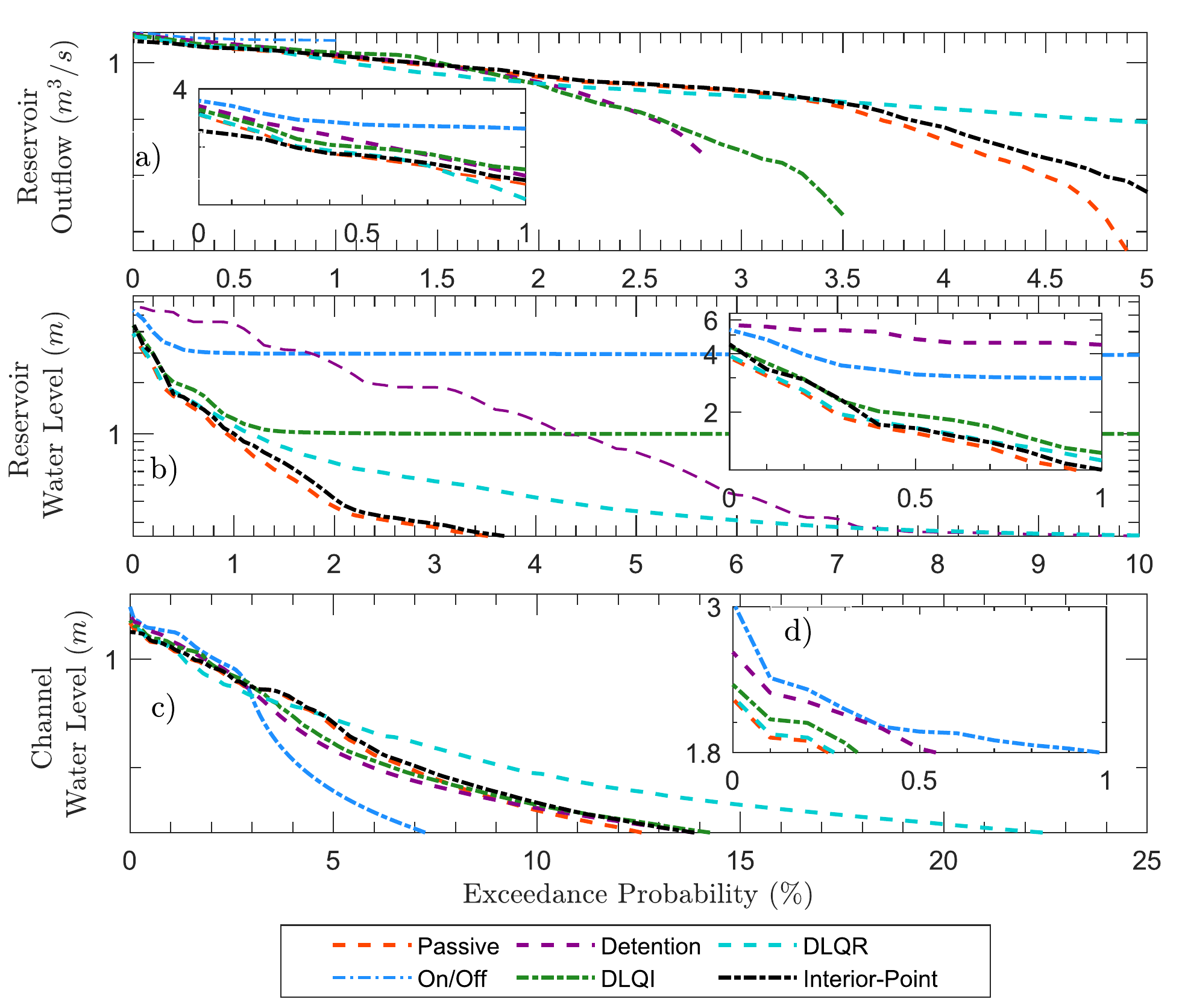}
    \caption{Duration curves comparing reactive and  predictive controls, where a), b), c), represent the exceedance probability of flows and levels in the reservoir, and levels in the channel, respectively, and d) represents a zoom of the channel flood exceedance probability}
    \label{fig:Frequency_Plots}
\end{figure*}

Frequency plots presented in Fig.~\ref{fig:Frequency_Plots} shows the exceedance probability of outflows and stages in the reservoir and channel. The on/off control is more likely to release high flows, which can cause erosion \cite{Schmitt2020}. This control is similar to a pond with the spillway in an elevation of 2.5 m above the ground. It certainly has the benefits of water quality enhancement due to larger retention times \cite{Sharior2019,Wong2018} but this approach was the one with higher outflows and hence water surface depths in the channel. The DLQR had the more spread flow-duration curve, indicating that this control typically can release flows throughout a larger period of time in a relatively small magnitude. This is particularly important since the pond can be slowly emptied when no future inflow is forecasted. The DLQI, as well as the on/off, had a flat duration curve around their regulated water surface depths of 3 m and 1 m, respectively, which indicates that they can satisfy their control goals over a rainy season.

Results shown in Fig.~\ref{fig:Frequency_Plots} part c) and d) show an estimate of the flood probability in the channel. They indicate that the passive control is the most suitable reactive control to avoid flood in the channel. However, the DLQR had approximately the same effect for mitigating the flood in the channel, but added the benefit of maintaining water in the channel and in the reservoir for longer periods of time. This type of RTC can, therefore, increase the period of time in the channel with some sort of baseflow \cite{xu2021enhancing}. The predictive control had no probability of flood in the channel, as desired from the optimization function.

Another interesting analysis is the trade-offs between the control efforts and the flood duration. The largest controls efforts were provided by the on/off, detention control and DLQI. Since we tune $Q$ with a high weight in the output deviation ($\dot{e} = r(k) - y(k)$), the control law rapidly tries to steer the system back to the defined reference setpoint, thus generating intense control efforts. This tuning can represent the operator preferences \cite{Troutman2020,fraternali2012putting} and could be a time-varying function in terms of the rainfall forecastings, although most of control algorithms assume a constant value. The control approach with least control effort was the predictive control, which nearly had 10\% of the detention control approach. The MPC problem for the continuous simulation was solved in approximately 30 minutes in a single core machine.

\section{Conclusions, Limitations, and Future Works} ~\label{sec:Conclusions}
The real-time control of watersheds, reservoirs and channels can decrease the risk of flooding in stormwater systems by better controlling their flow release over time. The application of different types of optimization-based and reactive controls algorithms was tested in a case study, and given the questions posed in Section \ref{sec:Model_Application}, the results support the following conclusions.

\begin{itemize}
    \item A1: The reactive controls have lower flood mitigation performance when compared to the predictive strategy, especially the water-quality controls of the on/off control and the detention control. 
    \item A2: The MPC control outperforms  all other control strategies for flood mitigation performance and also require less control effort.
    \item A3: The DLQR and DLQI are as effective as the passive control for flood mitigation in the channel. However, they add more specific benefits as maintain a desired water surface depth in the pond (DLQI) and increase residence times and low flows (DLQR). However, tuning the weighing matrices $\m Q$ and $\m R$ from the DLQR and LQR optimization problems can be complex due to different units in the objective function. Therefore, using normalized objective functions could be an alternative.
    \item A4: Despite the reactive controls were applied each 15-min and the predictive control each 60-min, the larger opportunity to change the valve operation had not compensate the lack of predictability of the states for the reactive controls, neither for large critical design storms events, nor for observed rainfall events.
\end{itemize}

Although only flood performance is evaluated in this paper, this type of modeling approach can be flexible enough to combine different control purposes according to the expected rainfall forecastings. Depending on the estimated future states of the system, one can change the objective functions of the system to shift from a flood focused strategy to a water quality control, increasing detention times. Moreover, since uncertainty in forecasting increase over the predicted horizon, it is possible to tune the weight matrices in the objective function following the uncertainty associated in the forecasting and thus giving more importance for short duration forecastings by increasing the cost weights associated with this duration.

Future application of the developed approach will consider scenarios of uncertainty in rainfall, model parameters, and measurements noise. Moreover, the approach needs to be tested in real case studies, such as the Upper San Antonio river watershed, which contains some of the most advanced flood protection systems in the US. Comparisons of the simplified dynamical system with the state of the art hydraulic and hydrologic models, such as SWMM, GSSHA, HEC-RAS, InfoWorks, and many others are also warranted. This methodology can be easily expanded for systems with many watersheds, reservoirs,
\section{Data Availability Statement}
Some or all data, models, or code generated or used during the
study are available in a repository or online in accordance with
funder data retention policies. All functions, scripts, and input data including the state-space non-linear model, the model predictive control algorithm, and the ruled-based algorithms are available in \cite{RTC_Code}.
\section*{Acknowledgment}
The authors gratefully acknowledge the support by the City of San Antonio, by the San Antonio River Authority, and by the National Science Foundation (NSF) under Grant 2015671.
\section{Supplemental Materials}
Tab. S1 is available online in an online repository found in \cite{GomesJr.2021}.



\bibliographystyle{IEEEtran}
\bibliography{References}

%




\end{document}

%% file: preambleT.tex
\usepackage{epsfig,color,amsmath}
 
\usepackage{wrapfig}
\usepackage{setspace}  
\usepackage{xcolor}
\usepackage{epstopdf}
\usepackage{amssymb}
\usepackage{url}
\usepackage{mathtools}
\usepackage{enumitem}
\usepackage{multirow}
\usepackage{makecell}
\usepackage{amsmath}
\usepackage{stackrel}
\usepackage{booktabs}
\usepackage[flushleft]{threeparttable}
\usepackage{makecell}

\usepackage{subfig}
\usepackage{graphicx}

\usepackage[linesnumbered,lined,boxed,commentsnumbered,ruled,longend]{algorithm2e}


\setlist[itemize]{leftmargin=*}

\DeclareMathOperator{\subjectto}{subject\ to}

\makeatother

\newcommand{\mat}[1]{\boldsymbol{#1}}

\providecommand{\mB}{\ensuremath{\mat{B}}}

\newcommand{\m}{\boldsymbol}
\allowdisplaybreaks[4]
\newcommand{\mc}[1]{\mathcal{#1}}
\newcommand{\mbb}[1]{\mathbb{#1}}
\newcommand{\mr}[1]{\mathrm{#1}}

\usepackage{amsthm}

\usepackage[usestackEOL]{stackengine}
\stackMath
\usepackage[framemethod=TikZ]{mdframed}
\usepackage{bm}
\mdfdefinestyle{MyFrame}{%
	linecolor=black,
	outerlinewidth=1.5pt,
	roundcorner=1.5pt,
	innerrightmargin=5pt,
	innerleftmargin=5pt,}

\usepackage{tikz}
\usetikzlibrary{matrix,positioning,decorations.pathreplacing}

\usepackage[export]{adjustbox}